\documentclass[11pt]{article}
\usepackage{mathrsfs}
\usepackage{amsmath}
\usepackage{amsfonts}
\usepackage{amssymb, amsmath, cite}
\usepackage{color}
\usepackage{graphicx}

\newtheorem{theorem}{Theorem}

\newcommand\be{\begin{equation}}
\newcommand\ee{\end{equation}}
\newcommand\ber{\begin{eqnarray}}
\newcommand\eer{\end{eqnarray}}
\newcommand\berr{\begin{eqnarray*}}
\newcommand\eerr{\end{eqnarray*}}
\newcommand\bea{\begin{eqnarray}}
\newcommand\eea{\end{eqnarray}}

\newcommand\dd{\mbox{d}}\newcommand\lm{\lambda}
\newcommand\e{\mathrm{e}}
\newcommand\pa{\partial}

\newcommand{\nn}{\nonumber}
\newcommand{\Erf}{\mbox{Erf}}

\newcommand{\Ei}{\mbox{Ei}}

\title{Determination of Angle of Light Deflection\\ in Higher-Derivative Gravity Theories}

\author{Chenmei Xu\\School of Mathematics and Statistics\\Henan University\\
Kaifeng, Henan 475004, PR China\\ \\
Yisong Yang\\Courant Institute of Mathematical Sciences\\ New York University\\New York, New York 10012, USA
}

\date{}

\begin{document}
\maketitle
\begin{abstract}
 {Gravitational light deflection is known as one of three classical tests of general relativity and the angle of deflection may be computed explicitly using approximate or exact solutions
describing the gravitational force generated from a point mass. In various generalized gravity theories, however, such explicit determination is often impossible due
to the difficulty with obtaining an exact expression for the deflection angle. In this work, we present some highly effective globally convergent iterative methods to determine the angle of semiclassical gravitational deflection 
 in higher- and infinite-derivative formalisms of quantum gravity theories. We also establish the universal properties that the deflection angle always
stays below the classical Einstein angle and is a strictly decreasing function of the incident photon energy, in these formalisms.

 }
\medskip

\medskip

\medskip


\end{abstract}

\section{Introduction}
\setcounter{equation}{0}
\setcounter{figure}{0}
\setcounter{table}{0}

Modification and generalization of the classical gravity theory of general relativity is an actively pursued rich subject in modern theoretical physics whose main motivations include attempt of a unification of
field theories, removal of singularities, handling of divergence, mechanism for matter accretion and galaxy formation, and the exponential expansion of universe driven by dark energy.
Among the numerous such extensions we may mention the Cartan--Einstein formalism, Brans--Dicke theory, conformal gravity,
Chern--Simons gravity, Kaluza--Klein type
theory, loop quantum gravity, and M-theory models. In order to examine the relevance of various extended theories, it would be desirable, if possible, to carry out some
appropriate comparisons with experimental or observable data. For general relativity, the three triumphantly well-known experimental tests are the gravitational  deflection of a light beam grazing the surface of Sun, the anomalous precession of the perihelion of Mercury's orbit, and the gravitational redshift of light, proposed by Einstein himself. Inevitably, these
tests and their associated calculations are of relevance and interest in the studies of the extended theories as well.
Due to the complicated behavior of gravitational interaction in general relativity
and its extended formulations, it is often a challenging task to calculate the deflection angle in full generality and such calculations may be categorized into two
types: explicit and implicit. Explicit calculations \cite{BH,BIS,BBD,BBH,BS,CM,E,EA,FS,IH,IP,MT,S,T,ZX} involve evaluating some complicated integrals often representable in terms of special functions (e.g., elliptic functions).
Implicit calculations \cite{ADGH,AGS,AHGH,AP,ALS,F,HHY,LC,Q,SB} amount to finding solutions to some complicated nonlinear equations. Of course, these two categories of problems are often related to or overlapped with each other
through further approximation and reduction.

 In this paper, we are interested in the determination of
the gravitational light-deflection angles arising in higher- and infinite-derivative  formalisms of quantum gravity which is accomplished through
solving
some nonlinear equations implicitly relating the deflection angles and various physical parameters in the extended theories to the classical Einstein angle.
The gravitational deflection problem studied here is semiclassical in the sense that the gravitational field is considered as a classical background applied field while the
incident photon
is regarded as a quantum particle scattering at the tree-level.

Recall that theory of higher-derivative gravity was initiated by Weyl \cite{W} shortly after Einstein's work on general relativity and aimed at a unified formalism of gravitation
and electromagnetism for which the extended Lagrangian contains higher powers of the Riemann tensor. Later this subject was revived by Stelle \cite{S1,S2} who
showed that suitably added quadratic terms of the Riemann tensor may render the quantum theory renormalizable. In cosmology, such extensions have helped
to enrich our understanding of inflation \cite{ALL,BG,BP,CYC,KDR} since the seminal work of Starobinsky \cite{Star}. On the other hand, however, massive spin-2 ghost
particles of negative probabilities inevitably arise in such a theory \cite{Lu}, a situation known as loss of unitarity. In order to overcome such a difficulty, infinite-derivative gravity theories \cite{Bis,Bis1,BMS,BMT,KMRS,Mod2,MR1,MR2,MR3,MT1} have successfully been developed. At fundamental levels,
the formalisms of higher- and infinite-derivative gravity theories have their natural origins in string theory \cite{CMod,FO,TBM}. It is in such profound context and
relevance that a series of analytic studies on the gravitational deflection
of photons in higher-derivative gravity theory \cite{ADGH,AGS,AHGH,AP} and infinite-derivative gravity
theory \cite{F} are carried out, aimed at obtaining bounds and estimates on the free parameters of the formalism, in which the key insight is contained in
the determination of deflection angles realized as the solutions of some nonlinear equations. Mathematically, these equations implicitly
but nonambiguously relate the deflection angles with various physical quantities, thus permitting an understanding of the dependence of the former on the latter, and
vice versa. In the present study we aim at achieving a systematic understanding of these important deflection-angle equations. We report our results in two areas
of interest,
computational and analytical: In the first area, we will see that, although the deflection angle equations in higher-derivative gravity theory and in infinite-derivative
gravity theory are of rather different technicalities, they share the common features that the solutions are uniquely determined by their physical parameters and
may be obtained by suitably designed globally convergent monotone iterative methods of the second order. These methods can be used effectively to determine
the deflection angles in all situations. In the second area we establish the dependence of the deflection angles on quantities such as the energy of an incident photon.
In particular, we prove that the deflection angle in the higher-derivative gravity theory \cite{ADGH,AGS,AHGH} and in the infinite-derivative gravity theory \cite{F} is always smaller than the classical Einstein angle. Note that such a result was first obtained in \cite{AHGH} in small-angle limit for the higher-derivative theory equation.
Here, however, we give proofs that this result is universally true in all situations and for the equations of full generality \cite{ADGH,AGS,AHGH,F}, among other results.

The content of the rest of the paper is as follows. In Section 2 we recall that the classical Einstein deflection angle, $\theta_{\mbox{E}}$, which is well known to be twice of the Newton
deflection angle, is not the exact value but an approximation of the deflection angle of a light beam grazing the surface of Sun, in both the weak-field approximation
and the Schwarzschild solution formalisms. Such a discussion helps with the contention that even in the simplest situation of general relativity the determination of
the deflection angle is by approximation but cannot be made with full accuracy. In Section 3 we follow \cite{ADGH,AGS,AHGH,AP,F} to present various
deflection angle equations to be solved to determine the deflection angle, $\theta$, arising in higher-derivative and infinite-derivative gravity theories.
In Section 4 we develop a series of monotone iterative methods to compute $\theta$ and establish the global convergence of the methods, meaning that our iterations
may be started from arbitrarily chosen initial states to achieve monotone convergence to the desired solutions. In Section 5 we present a collection of numerical examples using the computational methods obtained in the previous section. In Section 6 we analyze the deflection angle equations further and derive some
qualitative properties of the deflection angles in various situations. In particular, we show that $0<\theta<\theta_{\mbox{E}}$ is universally valid in
higher-derivative \cite{ADGH,AGS,AHGH} and infinite-derivative \cite{F} gravity theories, as was first obtained in \cite{AHGH} for the small-angle limit of
the higher-derivative deflection angle equation, although we also prove that this result is no longer valid for the deflection of a massive photon in the general relativity
gravity theory case studied in \cite{AP}. For example, we will see that $\theta$ may significantly exceed $\theta_{\mbox{E}}$ when the energy of the
incident massive photon is large.
In Section 7 we summarize our results.

\section{Einstein's gravitational deflection angle}
\setcounter{equation}{0}
\setcounter{figure}{0}
\setcounter{table}{0}

Consider a four-dimensional Minkowskian spacetime of the metric element $\dd s^2=g_{\mu\nu}\dd x^\mu\dd x^\nu$ of signature $(+---)$ governed by the
Einstein equation
\be
G_{\mu\nu}=-\frac{8\pi G}{c^4} T_{\mu\nu},\quad\mu,\nu=0,1,2,3,
\ee
where $G_{\mu\nu}=R_{\mu\nu}-\frac12 g_{\mu\nu}R$ is the Einstein tensor induced from the Ricci tensor $R_{\mu\nu}$ and Ricci scalar $R$, generated from
the spacetime metric tensor $g_{\mu\nu}$, $G$  the universal gravitational constant, $c$ the speed of light, and $T_{\mu\nu}$ the stress tensor of the matter
sector. We use
$\eta_{\mu\nu}=\mbox{diag}\{c^2,-1,-1,-1\}$ to denote the flat-spacetime Minkowski metric tensor. In the situation of a static point mass $M$ located at the
spatial origin ${\bf x}=(x^i)=(0,0,0)$, $T_{\mu\nu}$ is given by
\be
T_{\mu\nu}({\bf x})=c^2 M\eta_{\mu0}\eta_{\nu0}\delta({\bf x}),
\ee
where $\delta$ is the Dirac distribution, and the metric tensor is approximated by a weak-field formalism,
\be\label{2.3}
g_{\mu\nu}=\eta_{\mu\nu}+h_{\mu\nu},
\ee
with $h_{\mu\nu}$ a sufficiently weak field tensor which enables a linearization of the Einstein equation so that in the leading-order the metric element is
determined to be
\be
\dd s^2=\left(1-\frac{2GM}{c^2 r}\right) c^2\dd t^2-\left(1+\frac{2GM}{c^2 r}\right)\,\dd\ell^2,
\ee
with $r=|{\bf x}|=\sqrt{x^2+y^2+z^2}$ and $\dd\ell^2=\dd x^2+\dd y^2+\dd z^2$ being the Euclidean metric of the space. Thus the null condition $\dd s^2=0$
allows us to see that the speed of light varies as a function of $r$ following
\be\label{2.5}
v=\frac{\dd\ell}{\dd t}=c\left(\frac{1-\frac{\beta}{r}}{1+\frac{\beta}{r}}\right)^{\frac12},\quad \beta=\frac{2GM}{c^2},
\ee
which effectively defines a spatially dependent index of refraction, hence leading to the light bending. To see this in more detail, we consider a geometrically idealized
situation where the light beam is the closest to the point mass at $x=0,y=b>0,z=0$ and confined in the $xy$-plane. When $b$ is taken to be the radius of a spherically symmetric massive body so that one is interested in the bending of the light beam grazing the surface of the body, it is called the impact parameter. Then the angle of deflection for a photon
traveling from $x=-\infty$ to $x=\infty$ according to Einstein may be calculated to be
\be\label{2.6}
\theta=\frac1c\int_{-\infty}^\infty\left(\frac{\pa v}{\pa y}\right)_{y=b}\,\dd x.
\ee
An exact determination of the value of (\ref{2.6}) is impossible. However, for celestial bodies $\frac{\beta}r$ will be very small such that one may take the approximation
\be\label{2.7}
\frac1{1+\frac\beta r}\approx 1-\frac\beta r.
\ee
Inserting (\ref{2.7}) into (\ref{2.5}) we have
\be\label{2.8}
v\approx c\left(1-\frac\beta r\right).
\ee
Continue to assume that such an approximation remains valid under differentiation.  We obtain in the $xy$-plane
\be\label{2.9}
\frac{\pa v}{\pa y}\approx\frac{c\beta y}{(x^2+y^2)^{\frac32}}.
\ee
Substituting (\ref{2.9}) into (\ref{2.6}) we arrive at the  result of Einstein
\be\label{2.10}
\theta\approx \frac\beta b\int_{-\infty}^\infty\frac1{(\xi^2+1)^{\frac32}}\,\dd\xi=\frac{2\beta}b\equiv \theta_{\mbox{E}},\quad\xi=\frac xb,
\ee
where $\theta_{\mbox{E}}=\frac{2\beta}b$ is the celebrated Einstein angle, which is exactly twice of the Newton angle $\theta_{\mbox{N}}=\frac\beta b$ derived from
classical mechanics, as observed in the grazing light experiment around the sun.

One may question the legitimacy of the above approximation since the integration in (\ref{2.6}) is carried out over the full axis and error accumulation could become
intolerable. To settle such a question, we use (\ref{2.5}) in (\ref{2.6}) directly to get
\be\label{2.11}
\theta=\theta_{\mbox{N}}\int_{-\infty}^\infty h(\xi)\,\dd\xi,
\ee
where
\be
h(\xi)=\frac1{\sqrt{\xi^2+1}(\sqrt{\xi^2+1}-\theta_{\mbox{N}})^{\frac12}\left(\sqrt{\xi^2+1}+\theta_{\mbox{N}}\right)^{\frac32}}.
\ee
It is seen that the Einstein angle $\theta_{\mbox{E}}$ is recovered from (\ref{2.11}) in the limit $\theta_{\mbox{N}}=0$.
For the sun, we may use the values
\be\label{2.13}
G=6.674\times 10^{-11}\,\mbox{m}^3 (\mbox{kg})^{-1} \mbox{s}^{-2},\, M=2\times 10^{30} \,\mbox{kg},\, b=7\times 10^8\, \mbox{m}, \,c=3\times 10^8 \mbox{ms}^{-1}, 
%c=299792458\,\mbox{ms}^{-1},
\ee
resulting in $\theta_{\mbox{N}}=0.42374\times 10^{-5}$, and a high-precision numeric integrator with an accuracy of nine decimal places to obtain
\be\label{2.14}
\int_{-\infty}^\infty h(\xi)\,\dd\xi=1.999993344,
\ee
which is highly close to the magic number $2$ in the Einstein formula (\ref{2.10}).

One may also question the weak field approximation (\ref{2.3}) in the computation of the gravitational deflection angle and want to study the problem in terms of
an exact solution of the Einstein equation, say the Schwarzschild black hole metric of the form
\be\label{2.15}
\dd s^2=\left(1-\frac{GM}{2c^2 r}\right)^2\left(1+\frac{GM}{2c^2 r}\right)^{-2} c^2\dd t^2-\left(1+\frac{GM}{2c^2 r}\right)^4\,\dd\ell^2,
\ee
in isotropic coordinates \cite{Edd}. Thus, as before, the speed of light is given by
\be\label{2.16}
v=c\left(1-\frac{GM}{2c^2 r}\right)\left(1+\frac{GM}{2c^2 r}\right)^{-3}=16\, c\left(4-\frac\beta r\right)\left(4+\frac\beta r\right)^{-3},
\ee
which enjoys the identical first-order approximation (\ref{2.8}), hence, resulting in the Einstein angle $\theta_{\mbox{E}}$ again.
We may now compute (\ref{2.6}) in terms of the explicit formula (\ref{2.16}) to get
\be\label{2.17}
\theta=\theta_{\mbox{N}}\int_{-\infty}^\infty g(\xi)\,\dd\xi,
\ee
where the `weight' function $g(\xi)$ is given by
\be
g(\xi)=\frac{32\left(8\sqrt{\xi^2+1}-\theta_{\mbox{N}}\right)}{\left(4\sqrt{\xi^2+1}+\theta_{\mbox{N}}\right)^4},
\ee
which recovers (\ref{2.10}) in the limit $\theta_{\mbox{N}}=0$ as anticipated. For the data (\ref{2.13}) for the sun, we may carry out a hgh-precision numerical
integration of (\ref{2.17}) to find within nine decimal places the value
\be
\int_{-\infty}^\infty g(\xi)\,\dd\xi=1.999992512,
\ee
which is still close to the number $2$ of Einstein.

Consequently we have seen that the weak-field approximation and truncation up to the first order of the small-magnitude parameter $\theta_{\mbox{N}}=\frac\beta b=\frac{2GM}{c^2 b}$ offer us sufficiently accurate estimates for the gravitational deflection angle for realistic celestial bodies such as the sun. This point of view will be taken
in the subsequent analysis and computation.

\section{Nonlinear equations for the determination of semiclassical gravitational deflection angles in quantum gravity theories}
\setcounter{equation}{0}
\setcounter{figure}{0}
\setcounter{table}{0}

In this section, we briefly recall various nonlinear equations derived in literature for the determination of the deflection angles
arising in semiclassical gravitational deflection of photons for which gravity is taken as a classical field while the photon field is
quantized.
For convenience, we work in units with $c=1$ in subsequent discussion.

We start from the problem of the deflection of a massive photon of energy $E$ and mass $m$ by a gravitational
field \cite{AP} governed by the coupled action
\be\label{3.1}
S=\int\left(\frac R{16\pi G}-\frac14 B_{\mu\nu}B^{\mu\nu}+\frac12 m^2 W_\mu W^\mu\right)\,\sqrt{-g}\,\dd^4 x,
\ee
where $W^\mu$ is an Abelian vector boson field of mass $m$ and $B_{\mu\nu}=\pa_\mu W_\nu-\pa_\nu W_\mu$ the induced electromagnetic curvature tensor.
Using a weak field approximation for the equations of motion of the action (\ref{3.1}) and an unpolarized cross-section calculation
for the scattering of the Proca photon, it is shown \cite{AP} that the gravitational deflection angle $\theta$ of the incident photon and the Einstein angle $\theta_{\mbox{E}}$ are related through the implicit equation
\be\label{3.2}
\theta^2_{\mbox{E}}=\frac{1-\cos\theta}{\frac12\left(1+\frac{m^2}{2(E^2-m^2)}\right)^2+\frac13\left(1+\frac{m^2}{E^2-m^2}\right)(1-\cos\theta)\ln(1-\cos\theta)
-\frac1{12}(1-\cos\theta)^2},
\ee
whose small-angle approximation \cite{AP} is
\be\label{3.3}
\theta^2_{\mbox{E}}=\frac{\theta^2}{\left(1+\frac{m^2}{2(E^2-m^2)}\right)^2+\left(1+\frac{m^2}{E^2-m^2}\right)\frac{\theta^2}3\ln\frac{\theta^2}2}.
\ee

We note that, although this problem is not in the context of higher-derivative gravity theories, the deflection angle equations (\ref{3.2}) and (\ref{3.3}) provide interesting
and different
features in comparison with the features of the deflection angle equations of higher- and infinite-derivative gravity theories, as will be seen.

We next consider a superrenormalizable quantum gravity model of the nature of higher derivatives \cite{ALS,MS} governed by the simplified action \cite{AGS}
\be\label{3.4}
S=\int\left(\frac1{16\pi G} R+\frac\alpha2 R^2+\frac\beta2 R_{\mu\nu}R^{\mu\nu}+\frac A2 R\Box R +\frac B2 R_{\mu\nu}\Box R^{\mu\nu}\right)\sqrt{-g}\dd^4 x,
\ee
where the cosmological term is absent,  $\Box$ represents the d'Alembertian induced from $g_{\mu\nu}$, and $\alpha,\beta, A,B$ are coupling parameters.
Solving the linearized Einstein equation in the weak-field approximation
governing gravity around a point mass $M$ leads to the expressions \cite{MNS}
\be
h_{00}=MG\left(-\frac1r+\frac43 F_2-\frac13 F_0\right),\quad h_{11}=h_{22}=h_{33}=MG\left(-\frac1r+\frac23 F_2+\frac13 F_0\right),
\ee
for the nontrivial components of $h_{\mu\nu}$ where
\be
F_k=\frac{m^2_{k+}}{m^2_{k+}-m^2_{k-}}\frac{\e^{-m_{k-} r}}r+\frac{m^2_{k-}}{m^2_{k-}-m^2_{k+}}\frac{\e^{-m_{k+} r}}r,\quad k=0,2,
\ee
for the $k$-spin particles of the respective masses
\be
m^2_{2\pm}=\frac{\beta\pm\sqrt{\beta^2+\frac B{2\pi G}}}{2B},\quad m^2_{0\pm}=\frac{\sigma_1\pm\sqrt{\sigma_1^2-\frac{\sigma_2}{4\pi G}}}{2\sigma_2},\quad
\sigma_1=3\alpha+\beta,\quad \sigma_2=3A+B.
\ee

For such a solution, after working out the Feynman amplitude for the quantum scattering of a photon in the gravitational field, comparing the classical and tree-level cross-section formulas, and integrating the
unpolarized cross-section equation under the small-angle assumption, it is shown \cite{AGS}  that the gravitational deflection angle $\theta$ satisfies the equation
\bea\label{3.8}
\frac1{\theta^2_{\mbox{E}}}&=&\frac1{\theta^2}+\frac{E^2}{(m^2_{2-}-m^2_{2+})^2}\left(\frac{m^4_{2-}}{E^2\theta^2+m^2_{2+}}+\frac{m^4_{2+}}{E^2\theta^2+m^2_{2-}}\right)\nn\\
&&+\frac{2E^2}{m^2_{2-}-m^2_{2+}}\left(\frac{m^2_{2-}}{m^2_{2+}}\ln\left[\frac{E^2\theta^2}{E^2\theta^2+m^2_{2+}}\right]
-\frac{m^2_{2+}}{m^2_{2-}}\ln\left[\frac{E^2\theta^2}{E^2\theta^2+m^2_{2-}}\right]\right.\nn\\
&&\left.
-\frac{m^2_{2-} m^2_{2+}}{(m^2_{2-}-m^2_{2+})^2}\ln\left[\frac{E^2\theta^2+m^2_{2-}}{E^2\theta^2+m^2_{2+}}\right]\right),
\eea
which takes a much more complicated form than (\ref{3.2}).

In the extreme situation $m_{2-}\gg m_{2+}$, (\ref{3.8}) takes its approximate form \cite{AGS}
\be\label{3.9}
\frac1{\theta^2_{\mbox{E}}}=\frac1{\theta^2}+\frac1{\theta^2+\frac{m^2_{2+}}{E^2}}+\frac{2E^2}{m^2_{2+}}\ln\frac{\theta^2}{\theta^2+\frac{m^2_{2+}}
{E^2}},
\ee
which has also been derived in an earlier higher-derivative gravity theory context \cite{AA,AB,AHGH} and may be of independent interest. We will elaborate more on this link in Section 6 when we study some general properties of various semiclassical deflection angle equations.

Finally we recall that a rather drastic modification of the higher-derivative gravity theory (\ref{3.4}), first proposed by Modesto \cite{Mod}, is 
the infinite-derivative theory \cite{Bis,Bis1,BMS,BMT,Mod,MT1,MNS} defined by the action
\be
S=\frac1{16\pi G}\int\left(R+G_{\mu\nu}\left[\frac{a(\Box)-1}{\Box}\right] R^{\mu\nu}\right)\sqrt{-g}\,\dd^4 x,
\ee
where $a(\Box)=\e^{-\frac\Box{\Lambda^2}}$ is a differential operator of infinite order defined formally by the exponential power series expansion
of the Maclaurin type and $\Lambda$ a positive parameter measuring the non-locality scale of the theory. Solving the linearized Einstein equation in terms of the perturbed metric tensor $h_{\mu\nu}$, we have
\be\label{3.11}
h_{\mu\nu}=2GM\left(\frac{\eta_{\mu\nu}}r-\frac{2\eta_{\mu0}\eta_{\nu0}}r\right)\Erf\left(\frac{\Lambda r}2\right),
\ee
where $\Erf$ is the Gauss error function defined as
\be
\Erf(x)=\frac2{\sqrt{\pi}}\int_0^{x} \e^{-\xi^2}\,\dd\xi,
\ee
so that the Newton gravitational potential is given by
\be
\phi(r)=\frac12 h_{00}=-\frac{GM}r \Erf\left(\frac{\Lambda r}2\right),
\ee
which is singularity free at $r=0$. See also \cite{Ts} in a string theory context.
Consequently, in view of the method in \cite{AGS,AHGH,AP}, it is shown \cite{F}, by considering the Feynman diagram of the photon scattering
in the gravitational field described by (\ref{3.11}) and integrating the correspondingly deduced unpolarized cross-section equation, that the semiclassical gravitational deflection angle $\theta$ of a grazing photon of the total energy $E$
with the
impact parameter $b$ is the root of the equation
\be\label{3.14}
\frac1{\theta^2_{\mbox{E}}}=\frac1{\theta^2} \e^{-\frac{2\theta^2}{\lm^2}}+\frac2{\lm^2}\Ei\left(-\frac{2\theta^2}{\lm^2}\right),
\ee
where $\lm=\frac{\Lambda}E$ and $\Ei$ is the exponential integral function defined by
\be
\Ei(x)=-\int_{-x}^\infty\frac{\e^{-\xi}}\xi\,\dd\xi,
\ee
which is understood for $x>0$ in the sense of the Cauchy principal value but for our purpose we only consider $x<0$ which is classically defined.
It is well known that, using the Taylor expansion, one has the representation \cite{BO}
\be\label{3.15b}
\mbox{Ei}(x)=\gamma+\ln|x|+\sum_{k=1}^\infty\frac{x^k}{kk!},\quad x\neq0,
\ee
where $\gamma=0.57721566490...$ is Euler's constant, which may assist our computation in practice.

With $\sigma=\frac{2\theta^2}{\lm^2}$ and $\sigma_{\mbox{E}}=\frac{2\theta^2_{\mbox{E}}}{\lm^2}$, the equation (\ref{3.14}) becomes
\be\label{3.16}
\frac1{\sigma_{\mbox{E}}}=\frac{\e^{-\sigma}}\sigma +\Ei(-\sigma).
\ee
Consequently, if we truncate (\ref{3.15b}) in (\ref{3.16}) with a finite series, we may replace (\ref{3.16}) by its approximation
\be\label{3.18}
\frac1{\sigma_{\mbox{E}}}=\frac{\e^{-\sigma}}\sigma +\gamma+\ln\sigma+\sum_{k=1}^N(-1)^k\frac{\sigma^k}{kk!},\quad\sigma>0.
\ee

An immediate consequence of understanding the equations (\ref{3.2}), (\ref{3.3}), (\ref{3.8}), (\ref{3.9}), (\ref{3.14}), and (\ref{3.18}) is that it enables us to
see the dependence of the semiclassical gravitational deflection in various contexts on energy, a phenomenon known as dispersive deflection \cite{ADGH,AGS,AHGH,AP}.

\section{Methods for determination of semiclassical deflection angles and their global convergence}
\setcounter{equation}{0}
\setcounter{figure}{0}
\setcounter{table}{0}

For convenience, we rewrite (\ref{3.2}) as
\be\label{4.1}
\theta_{\mbox{E}}^2=\frac \tau{\frac18(1+B)^2+\frac13 B\tau\ln\tau-\frac1{12}\tau^2}\equiv f(\tau),
\ee
where $B=1+\frac{m^2}{E^2-m^2}$ and $\tau=1-\cos\theta$. Then we have $B>1$ and $0<\tau<1$ for our interest. Rewrite $f'(\tau)$ as
\be
f'(\tau)=\frac{g(\tau)}{\left(\frac18(1+B)^2+\frac13 B\tau\ln\tau-\frac1{12}\tau^2\right)^2}.
\ee
Then $g(1)=\frac1{24}(3B^2-2B+5)>0$ and $g'(\tau)<0$ for $0<\tau<1$. Thus $f'(\tau)>0$ ($0<\tau<1$). Since $f(0)=0$, we may assume
\be\label{4.3}
f(1)=\frac{24}{3(1+B)^2 -2}>\theta_{\mbox{E}}^2,
\ee
to ensure that (\ref{4.1}) has a unique solution.

In order to find the unique solution of (\ref{4.1}) constructively, we rewrite the equation as
\be\label{4.4}
\tau=\frac{\frac18(1+B)^2}{\frac1{\theta^2_{\mbox{E}}}-\frac13 B\ln\tau+\frac1{12}\tau}\equiv\varphi(\tau).
\ee
Then (\ref{4.3}) is identical to $\varphi(1)<1$. Since $\varphi(0^+)=0$ and $\varphi'(\tau)>0$ ($0<\tau<1$), we see that $0<\varphi(\tau)<1$ for $0<\tau<1$
which indicates the consistency of the fixed-point equation (\ref{4.4}). Moreover, pick any $\tau_0\in(0,1)$. Using induction and the monotonicity of $\varphi$, it can
be shown that, if $\tau_0<\varphi(\tau_0)$ (such a $\tau_0$ is called a subsolution of (\ref{4.4})), then the sequence $\{\tau_n\}$ defined by the iterative
scheme
\be\label{4.5}
\tau_{n+1}=\varphi(\tau_n)=\frac{\frac18(1+B)^2}{\frac1{\theta^2_{\mbox{E}}}-\frac13 B\ln\tau_n+\frac1{12}\tau_n},\quad n=0,1,2,...,
\ee
is strictly increasing; if $\tau_0>\varphi(\tau_0)$ (such a $\tau_0$ is called a supersolution of (\ref{4.4})), then the sequence $\{\tau_n\}$ defined by (\ref{4.5})
is strictly decreasing. In either situation, the sequence converges monotonically to the unique solution of (\ref{4.4}). In other words, we have established the
global convergence of the iterative scheme (\ref{4.5}) to the unique solution of (\ref{4.4}) or (\ref{4.1}) starting from any point $\tau_0\in(0,1)$.

We next consider (\ref{3.3}). With $\tau=\theta^2$, the equation (\ref{3.3}) becomes
\be\label{4.6}
\theta^2_{\mbox{E}}=\frac\tau{\frac14(1+B)^2+\frac13 B\tau\ln\frac\tau2}\equiv f_1(\tau).
\ee
We have
\be\label{4.7}
f_1'(\tau)=\frac{g_1(\tau)}{\left(\frac14(1+B)^2+\frac13 B\tau\ln\frac\tau2\right)^2},
\ee
with $g_1(\tau)=\frac14(1+B)^2-\frac13 B\tau$ which stays positive when $\tau<\frac32+\frac34\left(B+\frac1B\right)$ whose minimum for $B\geq1$ is attained at
$B=1$. Thus we are prompted to impose the range $0<\tau<3$ which suffices for our purposes since the deflection angle $\theta$ is small. Under such an
assumption the function $f_1$ is increasing and $f_1(0)=0$. Now assume
\be\label{4.8}
f_1(3)=\frac3{\frac14(1+B)^2+B\ln\frac32}>\theta^2_{\mbox{E}}.
\ee
Then it is ensured that (\ref{4.6}) has a unique solution, which will be constructed as follows.

As done earlier, rewrite (\ref{4.6}) as a fixed-point problem,
\be\label{4.9}
\tau=\frac{\frac14(1+B)^2}{\frac1{\theta^2_{\mbox{E}}}-\frac13 B\ln\frac\tau2}=\varphi_1(\tau).
\ee
Note that the denominator of (\ref{4.9}) stays positive in view of (\ref{4.8}). Besides, (\ref{4.8}) is equivalent to $\varphi_1(\tau)<3$ since $\varphi_1(\tau)$
increases for $0<\tau<3$. Consequently we arrive at the globally convergent monotone iterative scheme
\be\label{4.10}
\tau_{n+1}=\varphi_1(\tau_n) =\frac{\frac14(1+B)^2}{\frac1{\theta^2_{\mbox{E}}}-\frac13 B\ln\frac{\tau_n}2},\quad n=0,1,2,\dots,\quad \tau_0\in(0,3),
\ee
where the sequence $\{\tau_n\}$ increases or decreases according to whether $\tau_0<\varphi_1(\tau_0)$ or $\tau_0>\varphi_1(\tau_0)$.

Use $\psi$ to denote either $\varphi$ or $\varphi_1$ given in (\ref{4.4}) or (\ref{4.9}), respectively. Then $\psi'(\tau)>0$ in the respective interval of concern.
Thus, if $\tau_*$ denotes the unique fixed point of $\psi$ in the interval of concern and $\{\tau_n\}$ the sequence defined in either (\ref{4.5}) or (\ref{4.10}),
we have
\be
(\tau_*-\tau_{n+1})=\psi'(\xi)(\tau_*-\tau_n),\quad\mbox{ where }\xi\mbox{ lies between } \tau_*\mbox{ and }\tau_n,\quad n=1,2,\dots.
\ee
This estimate establishes that the rate of convergence of the sequence $\{\tau_n\}$ to $\tau_*$ is of the first order.

We note that the above developed sub- and supersolution method was adopted in \cite{XY} to solve the self-consistent energy gap equation \cite{KS,UCN,UN} for
 doped graphene 
superconductivity. Such an effective method grew out of a systematic treatment \cite{DY,Y1,Y2,Y3} of the Bardeen--Cooper--Schrieffer
gap equation in low-temperature
superconductivity theory.

We now consider (\ref{3.8}). For convenience, we set $\tau=E^2\theta^2, a=m^2_{2-}, b=m^2_{2+}, \lm=E^2$. Then (\ref{3.8}) becomes
\bea\label{4.11}
\frac1{\theta_{\mbox{E}}^2}&=&\frac \lm\tau+\frac \lm{(a-b)^2}\left(\frac{a^2}{\tau +b}+\frac{b^2}{\tau+a}\right)+\frac{2\lm}{a-b}\left(\frac ab\ln\frac{\tau}{\tau+b}
-\frac ba\ln\frac\tau{\tau+a}-\frac{ab}{(a-b)^2}\ln\frac{\tau+a}{\tau+b}\right)\nn\\
&\equiv &f_2(\tau).
\eea
Then $f_2(\tau)\to\infty$ as $\tau\to 0$ and $f_2(\tau)\to0$ as $\tau\to\infty$ so that (\ref{4.11}) always has a solution for some $\tau\in (0,\infty)$,
although it appears rather complicated.
However, we note that the first two derivatives of $f_2$ are nicely behaved:
\be\label{4.12}
f'_2(\tau)=-\frac{a^2b^2 \lm}{(\tau+a)^2(\tau+b)^2\tau^2},\quad f''_2(\tau)=\frac{2a^2 b^2 \lm(ab+2[a+b]\tau+3\tau^2)}{(\tau+a)^3(\tau+b)^3\tau^3}.
\ee
Hence (\ref{4.11}) has a unique solution in $(0,\infty)$, say $\tau_*$. We now proceed to construct $\tau_*$.

For the function $F_2(\tau)=f_2(\tau)-\frac1{\theta^2_{\mbox{E}}}$, we have $F_2(\tau_*)=0$. Take any $\tau_0>0$. From the Taylor expansion
\be
0=F_2(\tau_*)=F_2(\tau_0)+F'_2(\tau_0)(\tau_*-\tau_0)+\frac12 F''_2(\underline{\tau})(\tau_*-\tau_0)^2,
\ee
where $\underline{\tau}$ lies between $\tau_0$ and $\tau_*$, we obtain
\be\label{4.14}
\tau_*=\tau_0-\frac{F_2(\tau_0)}{F_2'(\tau_0)}-\frac{F''_2(\underline{\tau})}{2F'_2(\tau_0)}(\tau_*-\tau_0)^2.
\ee
Setting
\be
\tau_1=\tau_0-\frac{F_2(\tau_0)}{F_2'(\tau_0)},
\ee
we have in view of (\ref{4.14}) with the properties in (\ref{4.12})  that $\tau_*>\tau_1$ if $\tau_0\neq\tau_*$. Thus $0=F_2(\tau_*)<F_2(\tau_1)$, leading to
\be
\tau_2=\tau_1-\frac{F_2(\tau_1)}{F'_2(\tau_1)}>\tau_1.
\ee
Furthermore, replacing $\tau_0$ in (\ref{4.14}) by $\tau_1$, we get $\tau_*>\tau_2$. Consequently, after repeating these steps, we obtain
$\tau_1<\dots<\tau_n<\tau_*$ for any integer $n\geq1$ where
\be\label{4.17}
\tau_{n+1}=\tau_n-\frac{F_2(\tau_n)}{F_2'(\tau_n)},\quad n=0,1,2,\dots,
\ee
which is in fact the classical Newton iteration scheme, allowing us to get $\tau_*$ in the limit $n\to\infty$. From the above discussion, we see that the main advantage of our specific problem is that we can start the iteration from {\em any} initial state $\tau_0$. If $\tau_0\neq\tau_*$, we immediately go
below after the first iteration to obtain $\tau_1<\tau_*$. After this initial step, the sequence is ``tamed" to become monotone {\em increasing}, which
converges to the unique zero of $F_2(\tau)$ for $\tau>0$. In other words, we again have a globally convergent monotonically iterative method that allows
us to effectively construct or approximate the angle of deflection, $\theta$, given in (\ref{3.8}) or (\ref{4.11}).

We are now at a position to consider (\ref{3.9}). Set
\be
\lm^2=\frac{m_{2+}^2}{E^2}, \quad\theta^2=\lm^2\tau.
\ee
Then (\ref{3.9}) becomes
\be\label{4.19}
\frac{\lm^2}{\theta^2_{\mbox{E}}}=\frac1\tau+\frac1{1+\tau}+2\ln\frac\tau{1+\tau}\equiv f_3(\tau),\quad \tau>0.
\ee
Since $f_3(\infty)=0,f_3(0^+)=\infty$, the equation (\ref{4.19}) always has a solution $\tau>0$. Besides, we have
\be
f_3'(\tau)=  -\frac1{\tau^2(1+\tau)^2},\quad f_3''(\tau)=\frac{2(1+2\tau)}{\tau^3(1+\tau)^3},\quad \tau>0.
\ee
Thus the solution is unique and $f_3$ enjoys the same global properties as those of the function $f_2$ (being globally decreasing and concave up). As a consequence,
we conclude that, with the function $F_3(\tau)=f_3(\tau)-\frac{\lm^2}{\theta^2_{\mbox{E}}}$, we can invoke the iterative scheme
\be\label{4.21}
\tau_{n+1}=\tau_n-\frac{F_3(\tau_n)}{F'_3(\tau_n)},\quad n=0,1,2,\dots,
\ee
where $\tau_0>0$ is arbitrary and the sequence $\{\tau_n\}_{n\geq1}$ is monotone increasing and bounded  so that $\tau_*=\lim_{n\to\infty}\tau_n$ is the
unique solution of (\ref{4.19}).

Finally we solve (\ref{3.16}). We may rewrite it after an integration by parts as
\be\label{4.22}
\frac1{\sigma_{\mbox{E}}}=f_4(\sigma)=\frac{\e^{-\sigma}}\sigma+\e^{-\sigma}\ln\sigma-\int_\sigma^\infty \ln\xi\,\e^{-\xi}\,\dd\xi,\quad \sigma>0,\quad\sigma_{\mbox{E}}=\frac{2\theta^2_{\mbox{E}}}{\lm^2}.
\ee
It can be seen that $f_4(0^+)=\infty$ and $f_4(\infty)=0$ so that (\ref{4.22}) always has a solution. Furthermore, we have
\be\label{4.24}
f_4'(\sigma)=-\frac{\e^{-\sigma}}{\sigma^2}<0,\quad f''_4(\sigma)=\frac{\e^{-\sigma}}{\sigma^2}\left(1+\frac2\sigma\right)>0,\quad \sigma>0.
\ee
Thus the solution of (\ref{4.22}) is unique which can be approximated by the globally convergent monotone iterative scheme
\be\label{4.25}
\sigma_{n+1}=\sigma_n-\frac{F_4(\sigma_n)}{F'_4(\sigma_n)},\quad n=0,1,2,\dots,
\ee
where $F_4(\sigma)=f_4(\sigma)-\frac1{\sigma_{\mbox{E}}}$, $\sigma_0>0$ is arbitrary, the sequence $\{\sigma_n\}_{n\geq1}$ is monotone increasing, and
$\lim_{n\to\infty}\sigma_n=\sigma_*$, with $\sigma_*>0$ the unique solution of (\ref{4.22}).

We may investigate the rates of convergence of the iterative schemes (\ref{4.17}), (\ref{4.21}), and (\ref{4.25}). For convenience we use $\tau$ to replace the
variable $\sigma$ in (\ref{4.22}) and subsequent discussion, as in (\ref{4.11}), (\ref{4.19}), etc., and set
\be
\Psi(\tau)=\tau-\frac{F_i (\tau)}{F'_i(\tau)},\quad i=2,3,4,\quad\tau>0.
\ee
Then (\ref{4.17}), (\ref{4.21}), and (\ref{4.25}) are of the form $\tau_{n+1}=\Psi(\tau_n)$. Furthermore, we have
\be
\Psi'(\tau)=\frac{F_i(\tau)F_i''(\tau)}{(F_i'(\tau))^2},\quad i=2,3,4,\quad \tau>0.
\ee
Consequently, we have in particular $\Psi'(\tau_*)=0$ and
\be\label{4.28}
\Psi''(\tau_*)=\frac{F''_i(\tau_*)}{F_i'(\tau_*)}=\frac{f''_i(\tau_*)}{f'_i(\tau_*)}<0,\quad i=2,3,4.
\ee
Thus applying the Taylor expansion
\bea
\tau_{n+1}&=&\Psi(\tau_n)=\Psi(\tau_*)+\Psi'(\tau_*)(\tau_n-\tau_*)+\frac12\Psi''(\xi)(\tau_{n}-\tau_*)^2\nn\\
&=&\tau_*+\frac12\Psi''(\xi)(\tau_n-\tau_*)^2,
\eea
 with
$\xi$ lying between $\tau_*$ and $\tau_n$, we arrive at the estimate
\be
\tau_*-\tau_{n+1}=\frac12\Psi''(\xi)(\tau_*-\tau_n)^2=\mbox{O}((\tau_*-\tau_n)^2),
\ee
as $n\to\infty$, in view of (\ref{4.28}). In other words, the convergence $\tau_n\to\tau_*$ as $n\to\infty$ is of the rate of the exact second order.

For convenience we summarize our results of this section as follows.

\begin{theorem}
 Consider the semiclassical deflection angle equations (\ref{3.2}), (\ref{3.3}); (\ref{3.8}), (\ref{3.9}); (\ref{3.14}), arising in various quantum gravity theories.

\begin{enumerate}
\item[(i)] For (\ref{3.2}) and its small-angle approximation (\ref{3.3}), describing the deflection angle of a massive Proca photon propagating in the gravity field of general relativity, we may use the scheme (\ref{4.5}) and (\ref{4.10}), respectively, to construct iterative sequences which converge
monotonically to the unique solutions of the equations in the regimes of our interest. The rate of convergence is of the first order in either case.

\item[(ii)] For (\ref{3.8}) and its approximation (\ref{3.9}), determining the deflection angle of a photon in higher-derivative gravity theory, we may use
the scheme (\ref{4.17}) and (\ref{4.21}), respectively, to iteratively construct monotonically increasing sequences which converge to the unique solutions
of the equations in both cases. The convergence is global, meaning that it is independent of the choice of an initial state, and of the second order.

\item[(iii)] For (\ref{3.14}), of a non-local feature, determining the deflection angle of a photon in infinite-derivative gravity theory, we may use the
iterative scheme (\ref{4.25})
to obtain the unique solution of the equation, in the limit of the sequence. The sequence monotonically increases regardless of its initial state and converges with
a second-order rate.

\end{enumerate}
\end{theorem}

Before concluding this section, we discuss (\ref{3.18}) for practical purposes. To this end, we rewrite the equation as
\be\label{4.31}
\frac1{\sigma_{\mbox{E}}}=f_5(\sigma)=\frac{\e^{-\sigma}}\sigma+\gamma+\ln\sigma+\sum_{k=1}^N(-1)^k\frac{\sigma^k}{kk!},\quad\sigma>0.
\ee
We have $f_5(0^+)=\infty$ and
\be
f_5(1)=\e^{-1}+\gamma+\sum_{k=1}^N(-1)^k\frac1{kk!}<\e^{-1}+\gamma.
\ee
To ensure the existence of a solution of (\ref{4.31}) in $\sigma\in(0,1)$ (say), it suffices to assume
\be\label{4.33}
\e^{-1}+\gamma\leq\frac1{\sigma_{\mbox{E}}}.
\ee
However, this condition alone may not be enough to allow us to obtain a convergent iterative method. For the latter goal, we note the elementary facts
\be\label{4.34}
f_5'(\sigma)=-\frac1\sigma\left(\e^{-\sigma}+\frac{\e^{-\sigma}}\sigma-1\right)+\sum_{k=1}^N(-1)^k\frac{\sigma^{k-1}}{k!}<0,\quad \sigma\leq\ln 2,
\ee
\be\label{4.35}
f_5''(\sigma)=\frac{\e^{-\sigma}}\sigma+\frac2{\sigma^2}\left(\e^{-\sigma}+\frac{\e^{-\sigma}}\sigma-\frac12\right)+\sum_{k=2}^N(-1)^k\frac{\sigma^{k-2}}{k(k-2)!}>0,\quad \sigma\leq 1.
\ee
Thus, in order to be able to design a convergent iteration scheme as before following the properties (\ref{4.34}) and (\ref{4.35}), we need instead to work on a
smaller interval, say $(0,\ln 2)$, for $\sigma$. Consequently, we require $f_5(\ln 2)<\frac1{\sigma_{\mbox{E}}}$, which is guaranteed by the sufficient condition
\be\label{4.36}
\frac1{2\ln2}+\gamma\leq\frac1{\sigma_{\mbox{E}}},
\ee
which replaces (\ref{4.33}). In other words, when (\ref{4.36}) is satisfied, the equation (\ref{4.31}) has a solution in $\sigma\in(0,\ln2)$, which may be obtained
by the iterative scheme
\be\label{4.37}
\sigma_{n+1}=\sigma_n-\frac{F_5(\sigma_n)}{F'_5(\sigma_n)},\quad n=0,1,2,\dots,
\ee
where $\sigma_0\in(0,\ln2)$ and $F_5(\sigma)=f_5(\sigma)-\frac1{\sigma_{\mbox{E}}}$.  It should be noted that, unlike (\ref{4.25}), the convergence of the scheme (\ref{4.37}) is
local in the sense that it is valid only in a neighborhood of a solution of (\ref{4.31}).  It is also interesting that our iterative method is  applicable independent of $N$
although higher $N$ may provide a better approximation.

In the next section, we will present a series of numerical examples for the determination of the deflection angles in various models using the methods developed here.

\section{Numerical examples of deflection angles}
\setcounter{equation}{0}
\setcounter{figure}{0}
\setcounter{table}{0}

In this section we present a series of numerical computations of the deflection angles,
in various gravitational models discussed in Section 3, of a light beam grazing the surface of the sun with the data given in (\ref{2.13}), using the globally convergent
monotone iteration methods developed in Section 4.

We begin by considering the deflection angle equation (\ref{3.2}) for an incident photon of mass $m$ and energy $E>m$ which may be rewritten as (\ref{4.1}) with
a dimensionless parameter $B=1+\frac{m^2}{E^2-m^2}$ and $\tau=1-\cos\theta$ where $\theta$ is the deflection angle related back to $\tau$ by
\be\label{5.1}
\theta=2\arcsin\sqrt{\frac\tau2}.
\ee
For fixed $B>1$ and starting from any initial state $\tau_0>0$, we use the scheme (\ref{4.5}) to invoke an iterative sequence $\{\tau_n\}$ which converges to the unique solution of (\ref{4.4}), giving rise to our
desired deflection angle $\theta$ through (\ref{5.1}). For our purpose, since the deflection angle $\theta$ is a quantity in the order of $10^{-6}$ (radians) because
the Einstein angle assumes the value
\be
\theta_{\mbox{E}}=\frac{4 G M}{bc^2}=8.486658827544972\times 10^{-6} \quad\mbox{(radians)}=1.750499038748774 \quad\mbox{(arcseconds)},
\ee
we may effectively set
our stopping criterion at
\be\label{5.1a}
|\tau_n-\tau_{n-1}|<10^{-16}.
\ee
That is, when (\ref{5.1a}) is achieved, we terminate the iteration and accept $\tau_n$ as a computational solution of (\ref{4.4}).
\begin{figure}[h!]
   \centering
   \includegraphics[scale=0.8]{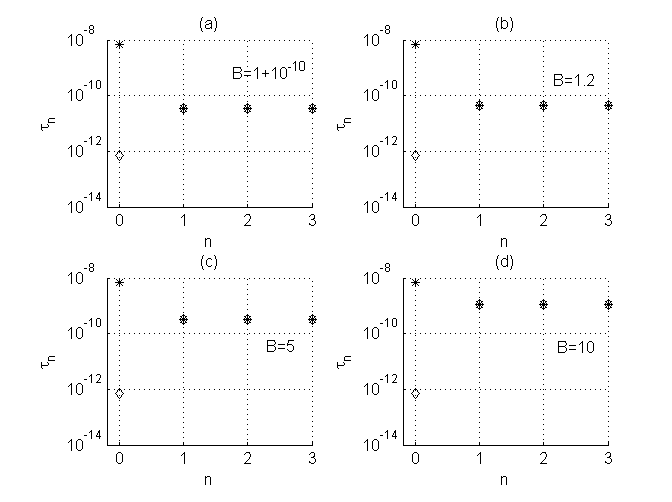}
   \caption{The plots of the iterative sequences defined by (\ref{4.5}) initiating at diverse initial states in the cases when the dimensionless parameter $B$ assumes values from near $1$ to 10. Although the initial states are far apart, the sequences converge quickly to yield the solutions of (\ref{4.1}) in a monotone manner after 3 iterations.}\label{F1}
\end{figure}

We choose $B=1+10^{-10}, 1.2, 5, 10$ as a variety of testing examples. Figure \ref{F1} shows the computed results. It is seen that in all examples the iteration stops
at $n=3$ indicating the effectiveness of the method. It is also seen that the sequence increases or decreases depending on whether $\tau_0<\varphi(\tau_0)$ or
$\tau_0>\varphi(\tau_0)$. Table \ref{T1} lists the corresponding results in terms of the computed deflection angle $\theta$. These results show that $\theta$
is uniformly bounded from below, monotonically increases as a function of $B>1$, and significantly exceeds $\theta_{\mbox{E}}$ for large
values of $B$. These facts will be rigorously established in Section 6.

\begin{table}[h]
%\caption{Nonlinear Model Results} % title of Table
\centering % used for centering table
\begin{tabular}{ccc} % centered columns (4 columns)
\hline%\hline %inserts double horizontal lines
$B$ & $\theta$ (in arcseconds)& ${\theta}/{\theta_{\mbox{E}}}$\\ [0.5ex] % inserts table
%heading
\hline % inserts single horizontal line
$1 +10^{-10}$& 1.750497801367231& 0.999999293126408\\ % inserting body of the table
1.2 & 1.925548955212175 & 1.100000007191391 \\
5& 5.251497002031886& 2.999999934753215 \\
10 & 9.627744776899915& 5.500000036436272\\[1ex] % [1ex] adds vertical space
\hline %inserts single line
\end{tabular}
% is used to refer this table in the text
\caption{Dependence of the deflection angle $\theta$ and the ratio of deflection angles $\frac{\theta}{\theta_{\mbox{E}}}$, determined by the equation (\ref{4.1}), with respect to the parameter $B$. The
results show that both are increasing functions of $B$, bounded uniformly from below, and assume large values at high levels of $B$.}\label{T1}  % title of Table
\end{table}

\begin{figure}[h]
   \centering
   \includegraphics[scale=0.8]{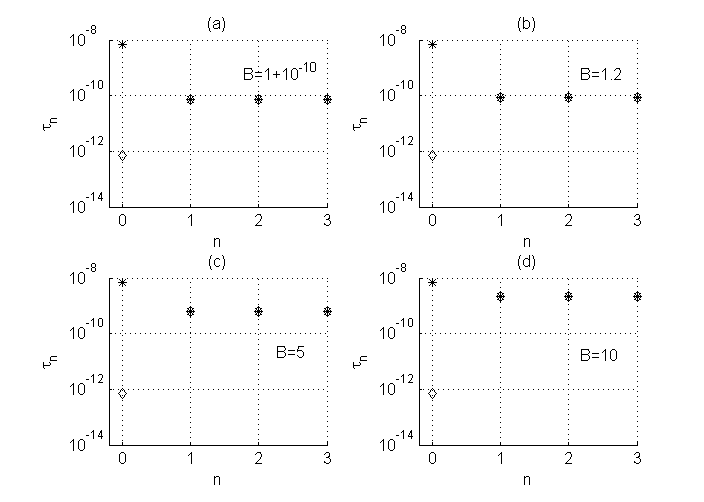}
   \caption{The plots of the iterative sequences defined by the scheme (\ref{4.10}) aimed at solving the simplified equation
(\ref{4.6}). The results show that the sequences converge globally to yield the solutions in a monotone manner after only 3  iterations which are satisfactorily close to
the solutions of the full equation (\ref{4.4}).}\label{F2}
\end{figure}

We continue to compute the solutions with $B=1+10^{-10},1.2, 5, 10$ for (\ref{4.6}) as a small-angle approximation for (\ref{4.4}),
using the iterative method (\ref{4.10}). Figure \ref{F2} shows the effectiveness of the method. The deflection angle is then determined through
%\be\label{5.2}
$\theta=\sqrt{\tau}.$
%\ee

Table \ref{T2} presents the corresponding results in terms of $\theta$. These are similar and in fact stay very close to those obtained from (\ref{4.4}). These results indicate
that (\ref{4.6}) is indeed an excellent approximation and simplification of (\ref{4.4}).

\begin{table}[h]
%\caption{Nonlinear Model Results} % title of Table
\centering % used for centering table
\begin{tabular}{ccc} % centered columns (4 columns)
\hline%\hline %inserts double horizontal lines
$B$ & $\theta$ (in arcseconds)& ${\theta}/{\theta_{\mbox{E}}}$\\ [0.5ex] % inserts table
%heading
\hline % inserts single horizontal line
$1 +10^{-10}$& 1.750499038331000& 0.9999999997613401 \\ % inserting body of the table
1.2 & 1.925548941961944& 1.099999999621990 \\
5& 5.251497109359388& 2.999999996065732 \\
10 & 9.627744689267194& 5.499999986374708\\[1ex] % [1ex] adds vertical space
\hline %inserts single line
\end{tabular}
% is used to refer this table in the text
\caption{Dependence of the deflection angle $\theta$ and the ratio of deflection angles $\frac{\theta}{\theta_{\mbox{E}}}$, determined
by the equation (\ref{4.6}), with respect to the parameter $B$. The
results are very close to and fluctuate around those obtained for the full equation (\ref{4.4}) in the whole  range of $B$ considered.}\label{T2}  % title of Table
\end{table}

\begin{figure}[h]
   \centering
   \includegraphics[scale=0.8]{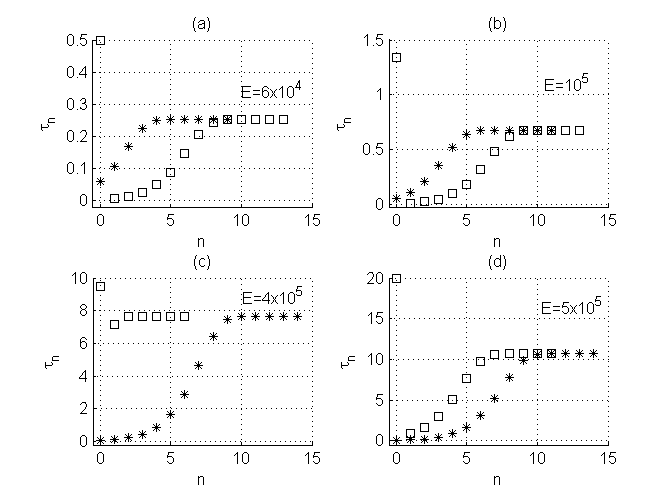}
   \caption{The plots of the iterative sequences constructed by the scheme (\ref{4.17}) aimed at solving (\ref{4.11}). The initial states are arbitrarily taken but the sequences become
monotone increasing after one iteration and converge quickly to yield the solutions after a little more than a dozen iterations.}\label{F4}
\end{figure}

We next consider (\ref{3.8}) or its reformulated form (\ref{4.11}). We choose the examples with $\beta=-6\times 10^6$ and $B=-20$ to fix
the parameters $a,b$ and  $E=6\times 10^4, 10^5, 4\times 10^5, 5\times 10^5$ as a series of tests for varied energy levels. For a variety
of initial states, the iterative sequence $\{\tau_n\}$ is
obtained using the scheme (\ref{4.17}) and plotted in Figure \ref{F4}, respectively, showing clearly the global convergence and monotonicity of sequence,
in each of the cases. In particular,
we see that
no matter how $\tau_0>0$ is chosen the sequence $\{\tau_n\}_{n\geq1}$ is always monotone increasing. The corresponding results for the deflection angle $\theta$
are presented in Table \ref{T3}. These results show that $\theta$ decreases as a function of the energy $E$ such that the ratio of deflections angles,$\frac\theta{\theta_{\mbox{E}}}$, always stays below 1.

\begin{table}[h]
%\caption{Nonlinear Model Results} % title of Table
\centering % used for centering table
\begin{tabular}{ccc} % centered columns (4 columns)
\hline%\hline %inserts double horizontal lines
$E$ & $\theta$ (in arcseconds)& ${\theta}/{\theta_{\mbox{E}}}$\\ [0.5ex] % inserts table
%heading
\hline % inserts single horizontal line
$6\times 10^4$& 1.725988097247495& 0.9859977406678280 \\ % inserting body of the table
$10^5$ & 1.696044917377635& 0.9688922300636846 \\
$4\times10^5$& 1.427391682449832& 0.8154198607673082 \\
$5\times 10^5$ & 1.350202399099214& 0.7713242733708186\\[1ex] % [1ex] adds vertical space
\hline %inserts single line
\end{tabular} % is used to refer this table in the text
\caption{Dependence of the deflection angle $\theta$ and the ratio of deflection angles $\frac{\theta}{\theta_{\mbox{E}}}$, determined by the equation (\ref{4.11}), with respect to the energy parameter $E$. The
results show that both are decreasing functions of $E$ and the latter always stays below 1 in the entire range of $E$.}
\label{T3} % title of Table
\end{table}

\begin{figure}[h]
   \centering
   \includegraphics[scale=0.8]{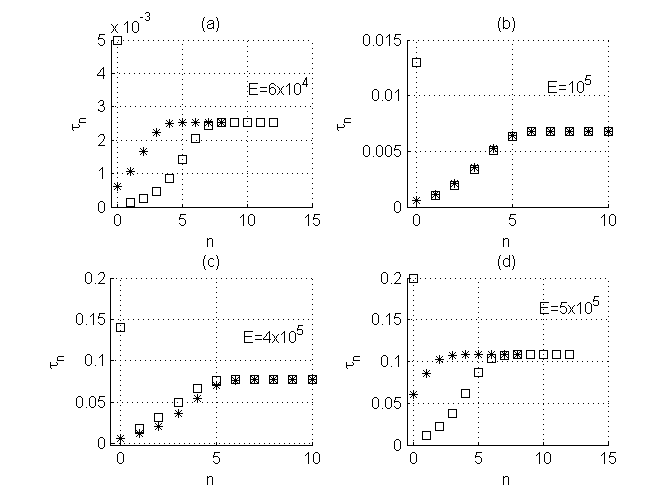}
   \caption{The plots of the iterative sequences obtained from the scheme (\ref{4.21}) designed to solve the simplified equation (\ref{4.19}). The
convergence is much faster and the results are close to the solutions to the full equation obtained in the previous example.}\label{F5}
\end{figure}

We then consider (\ref{3.9}) or its reformulated form (\ref{4.19}). In order to compare with the results obtained for (\ref{4.11}), we choose the same collection
of values of $E$. The iterative sequences constructed by the scheme (\ref{4.21}) originated from various initial states are plotted in Figure \ref{F5} so that the
corresponding results regarding the deflection angle $\theta$ are presented in Table \ref{T4}. Again these show that $\theta$ decreases with respect to $E$ and
$\frac\theta{\theta_{\mbox{E}}}$ stays forever below 1.

\newpage

\begin{table}[h]
%\caption{Nonlinear Model Results} % title of Table
\centering % used for centering table
\begin{tabular}{ccc} % centered columns (4 columns)
\hline%\hline %inserts double horizontal lines
$E$ & $\theta$ (in arcseconds)& ${\theta}/{\theta_{\mbox{E}}}$\\ [0.5ex] % inserts table
%heading
\hline % inserts single horizontal line
$6\times 10^4$& 1.725995361138118& 0.9860018902791454\\ % inserting body of the table
$10^5$ & 1.696060434897486& 0.9689010946900037\\
$4\times10^5$& 1.427463958247738& 0.8154611494491678 \\
$5\times 10^5$ & 1.350286859333170& 0.7713725226026583\\[1ex] % [1ex] adds vertical space
\hline %inserts single line
\end{tabular}
 % is used to refer this table in the text
\caption{Dependence of the deflection angle $\theta$ and the ratio of deflection angles $\frac{\theta}{\theta_{\mbox{E}}}$, determined by the equation (\ref{4.19})
which simplifies (\ref{4.11}), with respect to the parameter $E$. The
results are close to but slightly above those obtained for  (\ref{4.11}) which indicate that (\ref{4.19}) is a rather faithful but an over approximation of (\ref{4.11}) for the range of values of $E$
considered.}\label{T4} % title of Table
\end{table}

\begin{figure}[h]
   \centering
   \includegraphics[scale=0.76]{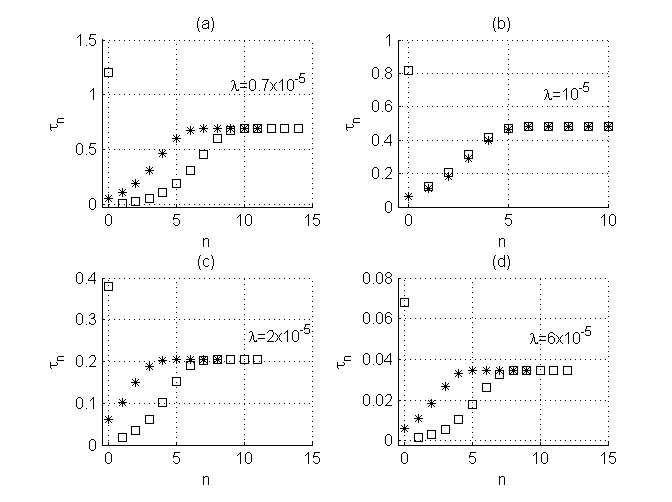}
   \caption{The plots of the iterative sequences obtained by the computational scheme (\ref{4.25}) aimed at solving the non-local equation (\ref{4.22})
governing the semiclassical gravitational deflection of a photon in infinite-derivative gravity theory. Global and monotone convergence is clearly demonstrated in all examples
regardless of the initial states. Fast convergence is exhibited by the termination of computation after about a dozen iterations.}\label{F7}
\end{figure}

Finally we study the determination of the deflection angle in the infinite-order derivative theory given by the equation (\ref{3.14}) or
its reformulated form (\ref{4.22}).
There is an extra technicality arising from handling an improper integral. To overcome this, we rewrite (\ref{3.16}) as
\be
\frac{\lm^2}{2\theta^2_{\mbox{E}}}=\frac1{\sigma_{\mbox{E}}}=\frac{\e^{-\sigma}}\sigma-\int_\sigma^1 \frac{\e^{-\xi}}{\xi}\,\dd\xi+\mbox{Ei}(-1), \quad \sigma>0,
\ee
 say, so that the improper part of $\mbox{Ei}(-\sigma)$ involving infinity upper bound for the integral is taken care of by a constant term. With this and the notation adopted for (\ref{4.22}), we work out the examples with $\lm=0.7\times 10^{-5}, 10^{-5}, 2\times 10^{-5}, 6\times 10^{-5}$,
respectively, originated from a variety of initial states. The iterative sequences constructed using the scheme (\ref{4.25}) are plotted in Figure \ref{F7}. The
global convergence and monotonicity of the method are again clearly exhibited in these plots. In Table \ref{T5} we present the results in terms of the deflection angle $\theta$. The monotone dependence of $\theta$ on $\lm$ is seen as well. As before, the ratio of the deflection angles, $\frac{\theta}{\theta_{\mbox{E}}}$ stays below 1.

\begin{table}[h]
%\caption{Nonlinear Model Results} % title of Table
\centering % used for centering table
\begin{tabular}{ccc} % centered columns (4 columns)
\hline%\hline %inserts double horizontal lines
$\lm$ & $\theta$ (in arcseconds)& ${\theta}/{\theta_{\mbox{E}}}$\\ [0.5ex] % inserts table
%heading
\hline % inserts single horizontal line
$0.7\times 10^{-5}$& 0.851487877726657& 0.4864257899480396\\ % inserting body of the table
$10^{-5}$ & 1.014487924079927& 0.5795421200602691 \\
$2\times10^{-5}$& 1.319831093291819& 0.7539741891176423 \\
$6\times 10^{-5}$ & 1.630892028807097& 0.9316726217529545\\[1ex] % [1ex] adds vertical space
\hline %inserts single line
\end{tabular}
% is used to refer this table in the text
\caption{Examples of the deflection angle $\theta$ and the ratio of deflection angles $\frac{\theta}{\theta_{\mbox{E}}}$, determined by the equation (\ref{4.22}), with respect to the parameter $\lm$. The
results show that both are increasing functions of $\lm$ and the latter stays below 1 for all values of $\lm$.}\label{T5}  % title of Table
\end{table}

%These properties will be proved in Section 6.

Interestingly all the properties (concerning the monotonicity and boundedness of the deflection angle) revealed in Tables \ref{T3}, \ref{T4}, and \ref{T5} have been deduced in \cite{AHGH} for (\ref{3.9}) in the small-angle equation limit for the semiclassical gravitational deflection of a photon in the context of the
fourth-order-derivative gravity theory \cite{S1,S2}. In the next section we shall
see that these properties are universally valid for all the equations considered here and can analytically be established.

\medskip

\begin{figure}[h]
   \centering
   \includegraphics[scale=0.8]{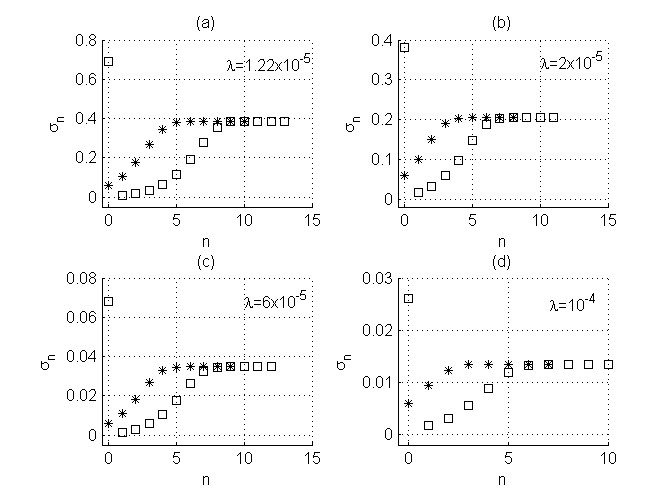}
   \caption{The behavior of iterative sequences computed by the scheme (\ref{4.37}) for the approximate equation (\ref{3.18}) with $N=2$. The starred and boxed plots represent results starting from two different initial states which are far apart. It is seen that after the initial iterations both sequences
are increasing and after a few iterations the sequences merge together, indicating
effective convergence into the desired solution, regardless of the initial states.}\label{F6}
\end{figure}

To end this section, we briefly discuss (\ref{3.18}) as an approximation to (\ref{3.14}). The locally convergent iterative method is given by (\ref{4.37}). It is well
checked that the condition (\ref{4.36}) is satisfied for the interest of our computation. The convergence for all tested $N=1,2,\dots$ follows the same pattern as that
for the full equation (\ref{3.14}): For any chosen initial state $\sigma_0$ the sequence $\{\sigma_n\}_{n\geq1}$ is monotone increasing and approaches its limiting
state, the desired solution, quickly. Figure \ref{F6} as a concrete example shows our computational results for $N=2$, which is a rather crude approximation of (\ref{3.14}). Nevertheless we see
that  convergence is clearly demonstrated and computation terminates after a dozen or some iterations.

The nature of the alternating series in (\ref{3.18}) gives rise to the interesting property that the solutions of (\ref{3.18})
associated with various integer values of $N$ are comparable. In fact, rewrite the right-hand side of (\ref{4.31}) as $f_{5,N}(\sigma)$. Then (\ref{4.34}) says that
$f'_{5,N}(\sigma)<0$ for all $0<\sigma\leq\ln2$ and $N=1,2,\dots$. Let $\sigma^N$ denote the solution of (\ref{4.31}) in $(0,\ln2)$. Then it is clear that, when
$N$ is an even integer, we have
\be\label{5.6}
f_{5,N+1}(\sigma)<f_{5,\infty}(\sigma)<f_{5,N}(\sigma),\quad\sigma\in (0,1).
\ee
Hence, since $\frac1{\sigma_{\mbox{E}}}=f_{5,N}(\sigma^N)=f_{5,\infty}(\sigma^{\infty})=f_{5,N+1}(\sigma^{N+1})$, we have in view of (\ref{5.6}) and the monotonicity of $f_{5,N}$ (for any $N$) the conclusion
\be\label{5.7}
\sigma^N>\sigma^{N+2}>\sigma^\infty>\sigma^{N+3}>\sigma^{N+1},\quad N=2,4,\dots.
\ee
In other words, an even (odd) integer $N$ in (\ref{3.18}) will serve to provide an upper (lower) estimate for the solution of the full equation (\ref{3.14}). Thus, naturally, we may use the average
\be
\overline{\sigma}=\frac{\sigma^N+\sigma^{N+1}}2
\ee
to approximate the solution of (\ref{3.14}) effectively, where $N$ is any integer. These results indicate that (\ref{3.18}) may be used effectively as a local equation to approximate the non-local equation (\ref{3.14}) and
that the solutions of (\ref{3.18}) with varying values of $N$ are well managed and controlled in the sense of (\ref{5.7}) that the results are comparable and may be
used to achieve an arbitrarily high accuracy in the limit.

\begin{figure}[h]
   \centering
   \includegraphics[scale=0.8]{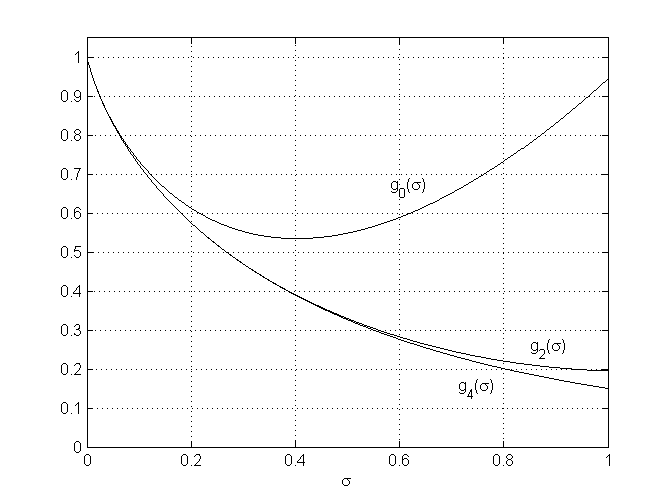}
   \caption{The plots of the functions $g_0(\sigma), g_2(\sigma)$, and $g_4(\sigma)$ for $\sigma\in (0,1]$. }\label{F3}
\end{figure}

The usefulness of (\ref{3.18}) prompts us to consider its solvability. For brevity and clarity, we will resort to the geometry of the equation which may be rewritten as
\be\label{5.9}
\frac{\sigma}{\sigma_{\mbox{E}}}=\sigma f_{5,N}(\sigma)\equiv g_N(\sigma).
\ee
Thus whether the equation has a solution is equivalent to whether the graph of the function $g_N(\sigma)$ has an intersection with the line of slope $\frac1{\sigma_{\mbox{E}}}=\frac{\lm^2}{2\theta^2_{\mbox{E}}}$ in the first quadrant. From the property (\ref{5.6}), we easily see that the equation always has
a solution when $N$ is odd. However, when $N$ is even, the situation is more subtle. For example, for $N=0,2,4$, we have
\be
g_0(\sigma)=\e^{-\sigma}+\gamma \sigma+\sigma\ln\sigma,\quad
g_2(\sigma)=g_0(\sigma)-\sigma^2+\frac{\sigma^3}4,\quad g_4(\sigma)=g_2(\sigma)-\frac{\sigma^4}{18}+\frac{\sigma^5}{96},
\ee
whose plots over $(0,1]$ are given in Figure \ref{F3}. First, it is seen that, since the curve of $g_0(\sigma)$
first descends and then ascends, (\ref{5.9}) with $N=0$ may fail to have a solution when $\frac1{\sigma_{\mbox{E}}}$ is small enough. In fact the minimum of $f_{5,0}(\sigma)=\frac{g_0(\sigma)}\sigma$ is
$0.9156891390$ which is attained at $\sigma=0.8064659942$. Hence (\ref{5.9}) with $N=0$ has no solution when $\frac1{\sigma_{\mbox{E}}}<0.9156891390$ or
$\lm<1.353284256\,\theta_{\mbox{E}}$. Next, the curve of $g_2(\sigma)$, descends to its lowest
level $g_2(1)=0.1950944412$ at $\sigma=1$, so that it cannot intersect a line through the origin with a slope lower than $0.1950944412$, indicating that
(\ref{5.9}) with $N=2$ has no solution when $\frac1{\sigma_{\mbox{E}}}<0.1950944412$ or $\lm<0.6246510085 \,\theta_{\mbox{E}}$.
Finally, as in the case of $g_2(\sigma)$,
the curve of $g_4(\sigma)$ descends to its lowest level $g_4(1)= 0.1499555523$ at $\sigma=1$. Hence (\ref{5.9}) with $N=4$ has no solution
when $\frac1{\sigma_{\mbox{E}}}<0.1499555523$ or $\lm<0.5476414015\,\theta_{\mbox{E}}$.
 These results suggest that, in order to approximate the non-local equation (\ref{3.14}) by the local one (\ref{3.18}) for
$\lm$ small, we need to maintain enough terms in the truncated series, which is naturally expected. Furthermore, from (\ref{5.9}), we have
\be\label{5.11}
\frac{\theta_N}{\theta_{\mbox{E}}}=\sqrt{g_N(\sigma^N)}, \quad\sigma^N\equiv\frac{2\theta_N^2}{\lm^2},\quad N=0,2,4,
\ee
where $\sigma=\sigma^N$ is the spot the curve of $g_N(\sigma)$ intersects the line $\frac\sigma{\sigma_{\mbox{E}}}$. Since $g_N(\sigma)<1$
in all cases, we are led to deduce
the consequence $\frac\theta{\theta_{E}}<\frac{\theta_N}{\theta_{\mbox{E}}}<1$ immediately, combining (\ref{5.7}) with (\ref{5.11}), where $\theta$ is the solution of the non-local equation (\ref{3.14}). This important result will be motivated further in Section 6 and rigorously proved there.

In Table \ref{T6}, we list our results computed with $N=2,5$ for (\ref{3.18})
and for the full equation (\ref{3.14}) (with $N=\infty$ in (\ref{3.18})). It is clear that for all the parameter values the deflection angles with $N=2$ are over estimates and
with $N=5$ lower estimates for the solutions of the full equation. Furthermore, although the truncation series are rather short (with $N=2,5$), the approximations are
already impressively accurate.

\begin{table}[h]
\centering
\begin{tabular}{cccc}
\hline
$\lm$ & $\theta$ (in arcseconds; $N=2$)&  $\theta$ (in arcseconds; $N=5$) & $\theta$ (in arcseconds; $N=\infty$)\\[0.5ex]
\hline
$1.22\times10^{-5}$ &1.106882965940261&1.105938485773018& 1.105938716591853\\
$2\times10^{-5}$&1.319907212427351 &1.319831090535894 &1.319831093291819\\
$6\times10^{-5}$&1.630892096617335 &1.630892028807040 &1.630892028807097\\
$10^{-4}$&1.693683591146825 &1.693683589574051 &1.693683589574069\\[1ex]
\hline
\end{tabular}
\caption{Examples of computed results using the approximate equation (\ref{3.18}) with $N=2,5$ in comparison with the solutions of the full equation (\ref{3.14})
corresponding to $N=\infty$ in (\ref{3.18}).}
\label{T6}
\end{table}

\section{Qualitative properties of deflection angles}
\setcounter{equation}{0}
\setcounter{figure}{0}

Following \cite{AHGH}, we consider Stelle's fourth-order-derivative gravity theory \cite{S1,S2} for which the gravitational action is taken to be
\be\label{6.1}
L=\int\left(\frac1{16\pi G}R+\frac\alpha2 R^2+\frac\beta2 R_{\mu\nu}R^{\mu\nu}\right)\sqrt{-g}\,\dd^4 x.
\ee
Under the assumption on the coupling parameters, $3\alpha+\beta>0$ and $\beta<0$, the masses of a spin 2 and a spin 0 particles of the model are given by
\be
m_2^2=-\frac1{8\pi G\beta},\quad m^2_0=\frac1{16\pi G(3\alpha+\beta)}.
\ee
Solving the linearized Einstein equations of the extended model (\ref{6.1}) subject to a centralized point mass $M$, with the stress tensor $T_{\mu\nu}({\bf x})
=M\eta_{\mu0}\eta_{\nu0}\delta^3({\bf x})$, it is found that the metric components corresponding to the three contributing terms in the action density
in (\ref{6.1}) are \cite{AHGH}
\bea\label{6.3}
h_{\mu\nu}^{(1)}&=&2GM\left(\frac{\eta_{\mu\nu}}r-\frac{2\eta_{\mu0}\eta_{\nu0}}r\right),\nn\\
h^{(2)}_{\mu\nu}&=&2GM\left(-\frac13\frac{\e^{-m_0 r}}r\eta_{\mu\nu}\right),\\
h^{(3)}_{\mu\nu}&=&2GM\left(-\frac23\frac{\e^{-m_2 r}}r\eta_{\mu\nu}+2\frac{\e^{-m_2 r}}r\eta_{\mu0}\eta_{\nu0}\right),\nn
\eea
respectively, so that the potential energy between two masses, $M_1$ and $M_2$, of a distance $r$, is of a leading Newton type supplemented by two additional
contributing terms
of a Yukuwa type,
\be
U(r)=M_1 M_2 G\left(-\frac1r-\frac13\frac{\e^{-m_0 r}}r+\frac43\frac{\e^{-m_2 r}}r\right).
\ee
Based on a calculation of momentum change of an incident photon coming from infinity and grazing the surface of Sun, it is shown \cite{AHGH} that the classical
deflection angle $\theta_{\mbox{C}}$ may be determined by the explicit formula
\be
\theta_{\mbox{C}}=\theta_{\mbox{E}}-{2GMb}\int_{-\infty}^\infty\frac{1+m_2 (\xi^2+b^2)^{\frac12}}{(\xi^2+b^2)^{\frac32}}\e^{-m_2(\xi^2+b^2)^{\frac12}}
\,\dd\xi.
\ee
In particular, we see that $\theta_{\mbox{C}}<\theta_{\mbox{E}}$, although it is not as transparent why $\theta_{\mbox{C}}>0$, which will not concern us here.

We now follow \cite{AHGH} to investigate the semiclassical gravitational deflection angle in the model.

Recall that, with (\ref{6.3}) and under the small-angle assumption, the relation between the Einstein deflection angle $\theta_{\mbox{E}}$ and that of the
semiclassical model, $\theta$, is
obtained \cite{AHGH} through a tree-level photon scattering calculation to be
\be\label{6.5}
\frac1{\theta^2_{\mbox{E}}}=\frac1{\theta^2}+\frac1{\lm^2+\theta^2}+\frac2{\lm^2}\ln\frac{\theta^2}{\lm^2+\theta^2},\quad\lm^2=\frac{m_2^2}{E^2}.
\ee
This equation is the same as (\ref{3.9}) or (\ref{4.19}). In \cite{AHGH}, it is deduced that, as a function of $\lm$,  the unique solution $\theta=\theta(\lm)$ of
(\ref{6.5}) has the properties $\theta(\lm)\to0, \lm\to0$, $\theta(\lm)\to\theta_{\mbox{E}},\lm\to\infty$, and $0\leq\theta(\lm)\leq \theta_{\mbox{E}}$ for all $\lm>0$.
Here, we first establish these results and the fact that $\theta(\lm)$ is strictly increasing. Thus, in particular, we have $0<\theta(\lm)<\theta_{\mbox{E}}$ for all $\lm>0$.
Subsequently, we generalize our analysis and show that these results are in fact universally valid for the full higher-derivative equation (\ref{3.8}) and the infinite-derivative equation (\ref{3.14}).

Indeed, with the function $f_3$ defined in (\ref{4.19}), we have
\be\label{6.6}
f_3'(\tau)\frac{\dd\tau}{\dd\lm}=-\frac1{\tau^2(1+\tau)^2}\frac{\dd\tau}{\dd\lm}=\frac{2\lm}{\theta^2_{\mbox{E}}}.
\ee
Thus $\tau(\lm)$ is a decreasing function of $\lm$ and $\tau(0^+)=\infty$ and $\tau(\infty)=0$. Furthermore, it can well be checked that the ratio
\be\label{6.7}
\rho(\lm)\equiv\frac{\theta^2}{\theta^2_{\mbox{E}}}=\tau(\lm)f_3(\tau(\lm))=1+\frac{\tau(\lm)}{1+\tau(\lm)}+2\tau(\lm)\ln\frac{\tau(\lm)}{1+\tau(\lm)}
\ee
satisfies
$
\lim_{\lm\to0^+}\rho(\lm)=0,\lim_{\lm\to\infty}\rho(\lm)=1,
$
since $\tau\ln\frac\tau{1+\tau}\to-1$ as $\tau\to\infty$. These properties are
as deduced in \cite{AHGH}. Moreover, we now show that $\rho(\lm)$ is monotone increasing. For this purpose, we rewrite (\ref{6.7}) as
\be\label{6.9}
\rho(\lm)=1+\tau(\lm) h(\tau(\lm)),\quad h(\tau)=\frac1{1+\tau}+2\ln\frac{\tau}{1+\tau},\quad \tau>0.
\ee
Then
%\be\label{6.10}
$
h'(\tau)=\frac{(2+\tau)}{\tau(1+\tau)^2}>0.
$
%\ee
Since $h(\infty)=0$, we get $h(\tau)<0$ for $\tau>0$. Furthermore
\be\label{6.11}
\rho'(\lm)=\tau'(\lm)\left(h(\tau(\lm))+\tau(\lm)h'(\tau(\lm))\right),
\ee
where
\be\label{6.11a}
h_1(\tau)=h(\tau)+\tau h'(\tau)=\frac2{1+\tau}+\frac1{(1+\tau)^2}+2\ln\frac\tau{1+\tau}
\ee
satisfies $h_1(\infty)=0$ and $h'_1(\tau)=\frac{2}{\tau(1+\tau)^3}$ ($\tau>0$). Therefore $h_1(\tau)<0$ for
all $\tau>0$ which implies $\rho'(\lm)>0$ for all $\lm>0$ in view of (\ref{6.11}). In particular, $0<\rho(\lm)<1$ or $\theta(\lm)<\theta_{\mbox{E}}$
for all $\lm>0$ and $\theta(0^+)=0$, $\theta(\infty)=\theta_{\mbox{E}}$. The numerical work in \cite{AHGH} shows that $\rho'(\lm)$ is a single-peak function.

Inserting (\ref{6.6}) and (\ref{6.11a}) into (\ref{6.11}), we obtain
\be\label{6.12}
\rho'(\lm)=-\frac{2\lm}{\theta^2_{\mbox{E}}}\tau^2\left(3+2\tau+2(1+\tau)^2\ln\frac\tau{1+\tau}\right),\quad\tau=\tau(\lm).
\ee

With (\ref{6.9}) and (\ref{6.12}), we may obtain the behavior of $\theta(\lm)$ and $\theta'(\lm)$ by
\be\label{theta}
\theta(\lm)=\theta_{\mbox{E}}\sqrt{\rho(\lm)},\quad\theta'(\lm)=\frac12\theta_{\mbox{E}}\frac{\rho'(\lm)}{\sqrt{\rho(\lm)}}.
\ee

We now consider the deflection angle problem \cite{F} in the infinite-order derivative gravity theory given in (\ref{3.16}), i.e.,
\be\label{6.13}
\frac{\lm^2}{2\theta^2_{\mbox{E}}}=f_4(\sigma)=\frac{\e^{-\sigma}}\sigma+\mbox{Ei}(-\sigma),\quad\sigma=\frac{2\theta^2}{\lm^2}.
\ee
From (\ref{4.22}) and (\ref{4.24}), we see that the unique solution, $\sigma(\lm)$, of (\ref{6.13}) satisfies
\be\label{6.14}
\sigma'(\lm)=-\frac{\lm}{\theta^2_{\mbox{E}}}\sigma^2(\lm)\e^{\sigma(\lm)}<0.
\ee
 Besides, we also have $\sigma(0^+)=\infty$ and
$\sigma(\infty)=0$. Furthermore we see that the ratio of deflection angles is given by
\be\label{6.15}
\rho(\lm)=\frac{\theta^2}{\theta^2_{\mbox{E}}}=\sigma f_4(\sigma)=\e^{-\sigma}\left(1+\sigma\ln\sigma\right)-\sigma\int_\sigma^\infty\ln\xi\,\e^{-\xi}\,\dd\xi,
\quad \sigma=\sigma(\lm).
\ee
Thus
$
\lim_{\lm\to0^+}\rho(\lm)=0,\lim_{\lm\to\infty}\rho(\lm)=1.
$
Moreover, since $f_4(\sigma)>0$ for any $\sigma>0$, we see in view of (\ref{3.16}), (\ref{4.22}), and (\ref{6.14}) that
\bea\label{6.16}
\rho'(\lm)&=&\sigma'(\lm)f_4(\sigma(\lm))+\sigma(\lm)f_4'(\sigma)\sigma'(\lm)
=\sigma'(\lm) \mbox{Ei}(-\sigma(\lm))\nn\\
&=&\frac\lm{\theta^2_{\mbox{E}}}\sigma^2(\lm)\e^{\sigma(\lm)}\int_{\sigma(\lm)}^\infty\frac{\e^{-\xi}}{\xi}\,\dd\xi>0.
\eea
In particular, $0<\rho(\lm)<1$ or $0<\theta(\lm)<\theta_{\mbox{E}}$ for all $\lm>0$, and $\theta(0^+)=0$, $\theta(\infty)=\theta_{\mbox{E}}$.
These properties are identical to those of the finite-order derivative model \cite{AHGH} discussed above. In particular, we expect the right-hand side of (\ref{6.16})
to be a single-peak function of $\lm>0$.

\begin{figure}[htb]
   \centering
   \includegraphics[scale=0.8]{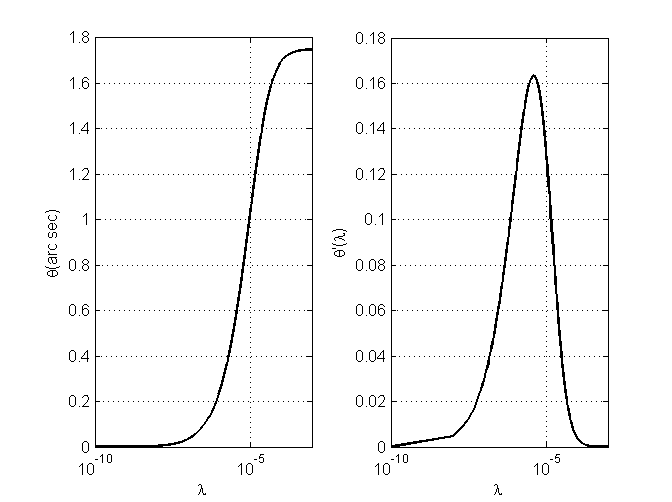}
   \caption{The plots of the deflection angle $\theta$ and its derivative, as functions of $\lm$, of the infinite-derivative gravity theory determined by the equation (\ref{3.14}), of a non-local feature  .}\label{F8}
\end{figure}

In view of (\ref{6.15}), (\ref{6.16}), and (\ref{theta}), we may obtain the behavior of $\theta$ and $\theta'$ with respect to $\lm$ as well.
Figure \ref{F8} shows the profiles of $\theta(\lm)$ and $\theta'(\lm)$.

With the afore-going study, we are now ready to investigate the seemingly more complicated equation (\ref{3.8}) or (\ref{4.11}) by the same method. For this
purpose, we rewrite (\ref{4.11}) as
\bea\label{6.17}
\rho(\lm)&=&\frac{\theta^2}{\theta_{\mbox{E}}^2}=\theta^2 f_2(\tau)\nn\\
&\equiv &1+f_6(\tau),\quad \tau=\tau(\lm),\quad\lm=E^2,
\eea
where
\be\label{6.20}
f_6(\tau)=\frac \tau{(a-b)^2}\left(\frac{a^2}{\tau +b}+\frac{b^2}{\tau+a}\right)+\frac{2\tau}{a-b}\left(\frac ab\ln\frac{\tau}{\tau+b}
-\frac ba\ln\frac\tau{\tau+a}-\frac{ab}{(a-b)^2}\ln\frac{\tau+a}{\tau+b}\right).
\ee

Rewrite (\ref{4.11}) as
\be\label{6.21}
\frac1{\theta^2_{\mbox{E}}}=F(\lm,\tau),\quad\lm=E^2.
\ee
Then, from (\ref{4.12}), we have $\frac{\pa F(\lm,\tau)}{\pa\lm}>0$ and $\frac{\pa F(\lm,\tau)}{\pa\tau}<0$ for $\lm>0,\tau>0$. Since, after
differentiating (\ref{6.21}), we have
%\be\label{6.21a}
$\frac{\pa F(\lm,\tau)}{\pa\lm}+\frac{\pa F(\lm,\tau)}{\pa\tau}\tau'(\lm)=0,
$
%\ee
we see that
 $\tau'(\lm)>0$ for all $\lm>0$.

From (\ref{6.17}) we have
%\be\label{6.22}
$\rho'(\lm)=f_6'(\tau)\tau'(\lm),$
%\ee
where $f_6'(\tau)$ takes a rather lengthy but elegant form
\bea
f_6'(\tau)&=&\frac1{(a-b)^2}\left(\frac{a^2}{\tau+b}+\frac{b^2}{\tau+a}\right)-\frac \tau{(a-b)^2}\left(\frac{a^2}{(\tau+b)^2}+\frac{b^2}{(\tau+a)^2}\right)\nn\\
&&+\frac2{a-b}\left(\frac ab\ln\frac\tau{\tau+b}-\frac ba\ln\frac\tau{\tau+a}-\frac{ab}{(a-b)^2}\ln\frac{\tau+a}{\tau+b}\right)\nn\\
&&+\frac{2\tau}{a-b}\left(\frac a{\tau(\tau+b)}-\frac b{\tau(\tau+a)}-\frac{ab}{(a-b)^2}\left[\frac1{\tau+a}-\frac1{\tau+b}\right]\right).
\eea
We can examine that $f_6'(\infty)=0$ and
$
f''_6(\tau)=\frac{2a^2 b^2(a+b+2\tau)}{\tau(\tau+a)^3(\tau+b)^3}>0,\tau>0.
$
Thus $f_6'(\tau)<0$ for all $\tau>0$ which leads us to arrive at $\rho'(\lm)<0$. Moreover, it may be checked directly that
%\be\label{6.25}
$\lim_{\tau\to 0^+}f_6(\tau)=0, \lim_{\tau\to\infty}f_6(\tau)=-1.$
%\ee
Besides, from (\ref{6.17}), it is seen that $\tau(\lm)\to 0$ as $\lm\to0$ and $\tau(\lm)\to\infty$ as $\lm\to\infty$. Consequently, in view of these
and (\ref{6.17}), we obtain the following consistent and summarized picture
\be
0<\rho(\lm)<1,\quad \rho'(\lm)<0,\quad\lm>0;\quad \lim_{\lm\to0^+}\rho(\lm)=1,\quad\lim_{\lm\to\infty}\rho(\lm)=0.
\ee

%As before, it will be interesting to study the quantity $\rho'(\lm)$ in terms of $\lm$. Note that (\ref{6.21a}) may be recast into the form
%\be\label{6.28}
%f_7(\tau)+\lm f_7'(\tau)\tau'=0,\quad \tau=\tau(\lm),
%\ee
%where
%\bea\label{6.29}
%f_7(\tau)&=&\frac1\tau+\frac 1{(a-b)^2}\left(\frac{a^2}{\tau +b}+\frac{b^2}{\tau+a}\right)\nn\\
%&&+\frac{2}{a-b}\left(\frac ab\ln\frac{\tau}{\tau+b}
%-\frac ba\ln\frac\tau{\tau+a}-\frac{ab}{(a-b)^2}\ln\frac{\tau+a}{\tau+b}\right),
%\eea
%which is a special case of the function $f_2$ with setting $A=1$ as defined in (\ref{4.11}) earlier. Substituting (\ref{6.28}) into (\ref{6.22}), we have
%\be\label{6.30}
%\rho'(\lm)=-\frac1\lm f'_6(\tau)\frac{f_7(\tau)}{f'_7(\tau)},\quad \tau=\tau(\lm),
%\ee
%where $f_6$ and $f_7$ are defined in (\ref{6.20}) and (\ref{6.29}), respectively, whose explicit expression is too complicated to put here. However, we derive again
%the result $\rho'(\lm)<0$ since $f'_6(\tau)<0$, $f_7(\tau)>0$, and $f_7'(\tau)<0$ for all $\tau>0$ in (\ref{6.30}).

We now study the semiclassical deflection angles determined by (\ref{4.1}) and (\ref{4.6}), respectively.
For simplicity, we consider (\ref{4.6}) first.

With $\theta^2=\tau=\tau(B)$ in (\ref{4.6}), we obtain by implicit differentiation the relation
\be\label{6.31}
f_1'(\tau)\tau'-\frac\tau{\left(\frac14(1+B)^2+\frac13B\tau\ln\frac\tau2\right)^2}\left(\frac12(1+B)+\frac13\tau\ln\frac\tau2\right)=0.
\ee
Since the minimum of the function $\tau\ln\frac\tau2$ is $-2\e^{-1}$ (attained at $\tau=2\e^{-1}$), so $\frac12(1+B)+\frac13\tau\ln\frac\tau2>0$
in the regime of our interest. Inserting this and (\ref{4.7}) into (\ref{6.31}), we arrive at the conclusion $\tau'(B)>0$. Consequently, (\ref{4.6}) indicates that
the ratio
\be\label{6.32}
\rho(B)=\frac{\theta^2}{\theta^2_{\mbox{E}}}=\frac{\tau(B)}{\theta^2_{\mbox{E}}}=\frac14(1+B)^2+\frac13 B\tau(B)\ln\frac{\tau(B)}2,
\ee
as a function of $B$, increases for $B>1$.
In particular, we find the uniform lower bound
\be\label{6.33}
\rho(B)>\rho(1)=1+\frac13\tau(1)\ln\frac{\tau(1)}2\geq 1-\frac23\e^{-1}\approx 0.7547470392,\quad B>1.
\ee
Thus $\rho(B)$ can never be made to assume arbitrarily small values. Besides, for given $B>1$, we have
\bea\label{6.34t}
\rho(B)&=&\frac14(1+B)^2+\frac13 B\tau(B)\ln\frac{\tau(B)}2\geq \frac14(1+B)^2-\frac23 B\e^{-1}\nn\\
&=&\frac14B^2+\frac12\left(1-\frac43\e^{-1}\right)B+\frac14.
\eea
Thus $\rho(B)$ may assume arbitrarily large values when $B$ is large. For example, the right-hand side of (\ref{6.34t}) indicates that
\be\label{6.35}
\rho(B)> \kappa\geq1\quad\mbox{(say)}\quad\mbox{when }B>\sqrt{\left(1-\frac43\e^{-1}\right)^2+(4\kappa-1)}-\left(1-\frac43e^{-1}\right).
\ee

Similarly, for (\ref{4.1}), we have
\be\label{6.33a}
f'(\tau)\tau'-\frac\tau{\left(\frac18(1+B)^2+\frac13B\tau\ln\tau-\frac1{12}\tau^2\right)^2}\left(\frac14(1+B)+\frac13\tau\ln\tau\right)=0.
\ee
Since the minimum of the function $\tau\ln\tau$ is $-\e^{-1}$ (attained at $\tau=\e^{-1}$) and $B>1$, we know that $\frac14 (1+B)+\frac13\tau\ln\tau>0$.
In view of this result and (\ref{6.33a}), we get again $\tau'(B)>0$. Thus the quantity
\be\label{6.34}
q(B)=\frac{2(1-\cos\theta)}{\theta^2_{\mbox{E}}}=\frac{2\tau(B)}{\theta^2_{\mbox{E}}}=\frac14(1+B)^2+\frac23 B\tau(B)\ln\tau(B)-\frac16\tau^2(B)
\ee
resembles the ratio of the angles of deflection and is an increasing function of $B>1$. As earlier in (\ref{6.33}), we have the uniform lower bound
\be
q(B)>q(1)\geq1-\frac23\e^{-1}-\frac16\approx 0.5880803725,\quad B>1,
\ee
since $0<\tau<1$. Furthermore, as in (\ref{6.34t}), we have
\be
q(B)\geq \frac14B^2+\frac12\left(1-\frac43\e^{-1}\right)B+\frac1{12},\quad B>1,
\ee
so that
\be\label{}
q(B)>\kappa\geq 1\quad\mbox{(say)}\quad\mbox{when }B\geq\sqrt{\left(1-\frac43\e^{-1}\right)^2+\left(4\kappa-\frac13\right)}-\left(1-\frac43e^{-1}\right).
\ee

It will be interesting to study the ratio of the deflection angles in its original setting, namely,
\be\label{6.41}
\rho(B)=\frac{\theta^2}{\theta^2_{\mbox{E}}}=q(B)\frac{\theta^2}{2\tau}.
\ee
For this quantity we have
\bea
\rho'(B)&=&q'(B)\frac{\theta^2}{2\tau}+q(B)\frac{\dd}{\dd\theta}\left(\frac{\theta^2}{2\tau}\right)\frac{\dd\theta}{\dd B}\nn\\
&=&\tau'(B)\left(\frac{\rho}{\tau}+\frac{\tau\omega}{\theta^2_{\mbox{E}}\sin\theta}\right)(B),
\eea
where
\be
\omega=\omega(\theta)=\frac{\dd}{\dd\theta}\left(\frac{\theta^2}\tau\right)=\frac{2\theta(1-\cos\theta)-\theta^2\sin\theta}{(1-\cos\theta)^2},
\ee
and we have used the fact $\theta'(B)=\frac1{\sin\theta} \tau'(B)$. It can be readily examined that $\omega(\theta)>0$ for $0<\theta<\frac\pi2$. This again establishes that $\rho'(B)>0$ for our problem.

Moreover, using $\omega(\theta)>0$ ($0<\theta<\frac\pi2$), we deduce the bounds
\be
\lim_{\theta\to0}\frac{\theta^2}{\tau(\theta)}=2<\frac{\theta^2}{\tau(\theta)}<\frac{\pi^2}4=\lim_{\theta\to\frac\pi2}\frac{\theta^2}{\tau(\theta)},\quad
0<\theta<\frac\pi2,
\ee
which may be used in conjunction with the estimates for $q(B)$ to obtain corresponding estimates for $\rho(B)$ for $B>1$ in view of (\ref{6.41}).

We may summarize our study of this section as follows.

\begin{theorem} Consider the light deflection angle $\theta$ determined implicitly by the equations (\ref{4.11}), (\ref{4.19}), and (\ref{4.22}),
which are the deflection-angle equations (\ref{3.8}), (\ref{3.9}), and (\ref{3.14}), in their original forms, arising in the higher-derivative
and infinite-derivative gravity theories, respectively, as a well-defined function of the total energy $E$ of the incident photon. In all these cases $\theta$ is a strictly decreasing function
of $E$ and satisfies the universal bounds 
\be
0<\theta<\theta_{\mbox{\rm E}},\quad E>0,
\ee
where $\theta_{\mbox{\rm E}}$ is the classical Einstein deflection angle, so that 
\be
\theta\to 0\quad\mbox{as }E\to\infty,\quad \theta\to\theta_{\mbox{\rm E}}\quad\mbox{as }E\to0.
\ee
However, such properties are not all valid for the deflection angle in the model of general-relativity
gravity coupled with a Proca photon field defined by the equations (\ref{3.2}) and (\ref{3.3}). More precisely, for these equations rewritten in the forms (\ref{4.1}) and
(\ref{4.6}), respectively, in terms of the dimensionless parameter $B=1+\frac{m^2}{E^2-m^2}$
given in terms of the mass $m$ and energy $E$ of an incident photon, the deflection angle $\theta$ strictly increases with respect to $B$, or decreases
with respect to $E$, but $\theta$ is bounded below uniformly away from zero for all $B$ and ${\theta}$ grows linearly when $B$ assumes large values so that $\theta$ exceeds $\theta_{\mbox{E}}$
arbitrarily in the limit $m\to E$.
\end{theorem}

In particular, for (\ref{4.1}) and (\ref{4.6}), we see that $\theta(B)$ exceeds $\theta_{\mbox{E}}$ when $B$ is sufficiently large and $\theta(B)\not\to0$ as $B\to1$,
corresponding to $E\to\infty$ or $m\to0$ in (\ref{3.2}) and (\ref{3.3}).

\section{Conclusions}

In this paper we have carried out a systematic study of the computational and analytical aspects aimed at the determination of the angle of light deflection in
the semiclassical settings of higher- and infinite-derivative formalisms of quantum gravity theories. As a result, we conclude with the following.

\begin{enumerate}
\item[(i)] For all the equations which define the implicit dependence of the angle of light deflection on various physical parameters, globally convergent
monotone iterative methods are developed. These methods have the common features that in their implementation the convergence is indifferent to the choice of
initial states and the convergence rate is of the second order.

\item[(ii)] A series of numerical examples are presented which demonstrate the effectiveness of the iterative methods for the determination of the angle of
deflection in each of the cases above. In fact, in all these examples, computation is mostly completed after a dozen or so iterations, even with a rather high-accuracy
termination threshold.

\item[(iii)] For the infinite-derivative gravity theory, the deflection-angle equation is non-local which adds complication in its practical handling. Based on a
finite-term-series truncation an approximation of the non-local equation is obtained which renders a local equation. It is shown that for the determination of
the deflection angle this local equation may be similarly solved by a locally convergent iterative method. More importantly, the properties of this local equation given by a
finite $N$-term series are well controlled in terms of $N$ so that, when $N$ is even the solution is an upper estimate and when $N$ is odd the solution is a lower estimate,
of the solution of the full non-local equation.

 Numerical examples are presented which show that
the solutions of the approximate equations are good approximations of the solution of the non-local equation.

\item[(iv)] In the case of a propagating photon governed by the Proca equations coupled with classical general relativity gravity, the deflection-angle equation
may be solved iteratively by a sub- and supersolution method. That the iterative sequence is increasing or decreasing depends on whether the initial state is
taken to be a sub- or supersolution of a fixed-point equation. For this method the rate of convergence is of the first order instead. Numerical examples are
presented for this problem as well.

\item[(v)] For all the deflection-angle equations, including the higher- and infinite-derivative gravity equations, the general relativity coupled with
the Proca electromagnetism equation,  and their small-angle simplifications, the angle of deflection is always a monotone decreasing function of the
incident photon energy.

\item[(vi)] The deflection angle arising in higher- and infinite-derivative formalisms of quantum gravity theories does not exceed the classical Einstein deflection angle and tends to zero as the energy of the incident photon goes to infinity.
However, such a universal property is not valid for the deflection angle in the case of the propagation of a Proca photon subject to general relativity gravity.

\end{enumerate}

\medskip
\medskip

{\small{Yang was partially supported
by National Natural Science Foundation of China under Grant No. 11471100.}}

%\newpage

\end{document}